\documentclass{jfm_draft}

\usepackage[utf8]{inputenc}
\usepackage{graphicx}
\usepackage{overpic}
\usepackage[sort,numbers]{natbib}
\usepackage{mathtools}
\usepackage{amssymb}
\usepackage{amsbsy}
\usepackage{bm}
\usepackage{color}
\usepackage{tikz}
\usepackage[cal=euler]{mathalpha}
\usepackage{xcolor} 
\usepackage[hidelinks]{hyperref}
\usepackage[nameinlink,noabbrev]{cleveref}
\usepackage{siunitx}

\usepackage{my_macros}

\allowdisplaybreaks

\title{Generalised Jeffery's equations for rapidly spinning particles. Part 1: Spheroids}
\date{\today}

\shorttitle{Part 1: Effective dynamics of rapidly spinning spheroids in shear flow}
\shortauthor{Dalwadi, Moreau, Gaffney, Ishimoto, and Walker}
\author{M. P. Dalwadi\aff{1\corresp{\email{m.dalwadi@ucl.ac.uk}}}, C. Moreau\aff{2}, E. A. Gaffney\aff{3}, K. Ishimoto\aff{2}, B. J. Walker\aff{1,4}}

\affiliation{\aff{1}Department of Mathematics, University College London, London, WC1H 0AY, UK
\aff{2}Research Institute for Mathematical Sciences, Kyoto University, Kyoto, 606-8502, Japan
\aff{3}Wolfson Centre for Mathematical Biology, Mathematical Institute, University of Oxford, Oxford, OX2 6GG, UK
\aff{4}Department of Mathematical Sciences, University of Bath, Bath, BA2 7AY, UK}

\date{}

\begin{document}

\maketitle

\section*{Abstract}
The observed behaviour of passive objects in simple flows can be surprisingly intricate, and is complicated further by object activity. Inspired by the motility of bacterial swimmers, in this two-part study we examine the three-dimensional motion of rigid active particles in shear Stokes flow, focusing on bodies that induce rapid rotation as part of their activity. Here, in Part 1, we develop a multiscale framework to investigate these emergent dynamics and apply it to simple spheroidal objects. In Part 2 \citep{dalwadi2023emergentPart2}, we apply our framework to understand the emergent dynamics of more complex shapes; helicoidal objects with chirality. Via a multiple-scales asymptotic analysis for nonlinear systems, we systematically derive emergent equations of motion for long-term trajectories that explicitly account for the strong (leading-order) effects of fast spinning. Supported by numerical examples, we constructively link these effective dynamics to the well-known Jeffery's orbits for passive spheroids, deriving an explicit closed-form expression for the effective shape of the active particle, broadening the scope of Jeffery's seminal study to spinning spheroids.

\section{Introduction}
\label{sec: Introduction}
The dynamics of objects in fluid flow can be incredibly intricate. Even restricting consideration to a simple shear flow in the Stokes regime, which arises as a key component in the local linearisation of general laminar flows, the corresponding equations of motion can be difficult to analyze explicitly. However, there are certain classes of object for which analytic solutions are known, beginning with the celebrated work of \citet{Jeffery1922}. \citeauthor{Jeffery1922}'s seminal study considered the orientation of ellipsoids in a shear flow, deriving an exact dynamical system of three governing equations for a general ellipsoid. Moreover, for a spheroid (also referred to as an ellipsoid of revolution), the governing equations simplify to what are now known as \emph{Jeffery's equations}, for which exact solutions can be found explicitly. These solutions were shown to be in agreement with experiments by \citet{Taylor1923a} one year later.

Given the impact of the contribution by \citet{Jeffery1922}, the solutions to the derived system of equations are known as \emph{Jeffery's orbits}. Furthermore, these governing equations apply far beyond the stated geometric restrictions of \citeauthor{Jeffery1922}'s original study. In particular, the work of \citet{Bretherton1962} and \citet{Brenner1964a} showed that \citeauthor{Jeffery1922}'s equations also hold for the motion of general axisymmetric objects, with the equations of angular motion reducing simply to those that describe the evolution of a spheroid in flow. That is, the rotational dynamics of general axisymmetric objects behave as effective spheroids in shear flow, with a corresponding effective aspect ratio that depends on the particular object.

In all the work discussed above, the bodies are passive. However, many bodies of interest are active, such as living microswimmers or externally driven particles. The consideration of active bodies contributes additional propulsion and rotation to the dynamics, resulting in behaviours that are significantly more intricate than their passive counterparts \citep{wittkowski2012self,junot2019swimming,fung2022local}. Active matter has been the subject of much recent research, including studies of bacterial locomotion across scales \citep{hyon2012wiggling, constantino2016helical, lauga2016bacterial, aranson2022bacterial}, the swimming behaviours of spermatozoon \citep{gaffney2021modelling, gong2021reconstruction}, and in active suspensions more generally \citep{saintillan2013active,saintillan2015theory,saintillan2008instabilities}. Driven by advances in micromanipulation and microfabrication, there has also been theoretical and experimental study of externally driven artificial active matter, such as the rigid, magnetic helices explored by \cite{ghosh2009controlled,zhang2009artificial} (see also \citet{zhou2021magnetically} for a recent review).
In many active swimmer systems, including both biological and artificial active matter, motion often proceeds over separated timescales. Rapid movements (often oscillatory) dominate small-scale dynamics over shorter timescales, and these give way to larger scale, emergent trajectories over longer timescales, as may be readily inferred by inspecting the length and timescales of observed swimmer trajectories. Some examples include the trajectories of {\it Crithidia deanei} in  Fig.~1B of \citet{gadelha2007}; mouse sperm in  Fig.~13a of \citet{woolley2003a}; a magnetically driven bacterium-like artificial swimmer in Fig.~4d of \citet{huang2016}; and 
bacteria in Fig.~S3 of \citet{turner2016}. In particular, while bacteria flagellum rotation rates can vary extensively with viscosity, number of flagella and other factors, some bacteria exhibit more than  $10^3$ flagellum rotations per second, as reviewed by \cite{berg2000}. One example is smooth swimming monotrichous {\it V. alginolyticus} mutants lacking lateral flagella \citep{magariyama}, where the longer length- and time-scale dynamics of smooth swimming emerges from rapid rotation. 
Furthermore, theoretical studies also highlight the emergence of larger-scale trajectories from rapid small-scale movements, for example the direct numerical simulations of \citet{hyon2012wiggling} and \citet{park2019b} for bacteria. 

Recently, examples of such multiscale dynamics have been explored using the asymptotic method of multiple scales \citep{hinch_1991,Bender1999}, yielding systematically simplified dynamical systems that govern the dominant behaviours of objects and swimmers in flow \citep{Walker2022a,gaffney2022canonical,ma2022reaching}. Motivated by the recent successes of the method of multiple scales in different swimmer contexts, which have been predominantly limited to motion in two dimensions (with the notable exception of the three-dimensional model in the Appendix of \citet{ma2022reaching}), the primary aim of this study is to exploit separated timescales in a three-dimensional (3D) model of an active particle to derive effective governing equations. 
Here, drawing inspiration from a general class of rapidly spinning bodies that include bacteria-like swimmers, we investigate the dynamics of objects whose self-generated spinning motion is fast compared to the flow and self-generated translation timescales in the swimming problem. In particular, we will focus on objects whose geometry can be reasonably approximated as constant in time, noting that this assumption is often a feature of externally driven rigid objects and simple models of shape-changing swimmers \citep{Lauga2009,Lauga2020}. This approach is also consistent with the popular squirmer model, in which a self-generated slip velocity drives the motion of a rigid object \citep{Ito2019SwimmingModel,Magar2005,Ishimoto2013}.

Seeking generality, we abstract away from any particular application or interpretation to develop an appropriate general framework in Part 1, before using this to study spheroidal objects. In Part 2 \citep{dalwadi2023emergentPart2}, we apply the framework we develop in Part 1 to study more general-shaped objects (chiral particles with helicoidal symmetry). Hence, in Part 1, we are essentially investigating active counterparts of the passive spheroids in shear flow analysed by \citet{Jeffery1922} to understand the effect of self-driven translation and fast rotation on their dynamics. 
In order to retain tractability, we study particles with minimally complex active characteristics. Specifically, we assume that our objects generate a prescribed constant linear velocity and fast angular velocity, each fixed in a reference frame that evolves with the object orientation. Even with this minimal additional complexity, the governing equations we derive are a strongly nonlinear coupled system that requires a nonstandard and technically intricate multiple scales analysis. In particular, we must solve a nonlinear coupled 3D system at leading-order and, using the method of multiple scales for systems (see, for example, pp.~127--128 of \citet{dalwadi2014flow} or p.~22 of \citet{dalwadi2018effect}), a non-self-adjoint coupled 3D system at next-order.

Nevertheless, we are able to make significant analytic progress to derive emergent equations of motion that systematically account for the rapid spinning. We show that there is a natural and fundamental correspondence between these emergent equations and the classic equations of \citet{Jeffery1922}. Specifically, in terms of quantities that we derive analytically, we show that the emergent equations of motion are equivalent to Jeffery's equations of motion in appropriately transformed variables, with effective coefficients that describe the effective hydrodynamic spheroidal shape in terms of the original shape parameters and the rapid rotation.

To summarise, in this first part of our two-part study, we develop the multiscale theoretical framework to consider a rapidly spinning self-propelled particle in Stokes flow, and apply it to the specific problem of a spheroidal particle in 3D shear flow. We start by formulating the equations of motion for such a swimmer in \S \ref{sec: Equations of motion}. To introduce the multiple scales notation for the full system, we first analyse a more intuitive limit of the full system in \S \ref{sec: Bacterial sublimit}, limiting the axis of rapid spinning to the symmetry axis of the spheroid. This also allows us to illustrate the fundamental steps involved in a standard multiple scales analysis before using an appropriate modification for systems to analyse the full problem in \S \ref{sec: General emergent behaviour}, which includes general off-axis rapid spinning. We then summarise and discuss the key physical implications and conclusions of our technical analysis in \S \ref{sec: results} in a non-technical manner, for the benefit of those that wish to skip the details of the analysis. Finally, we discuss our results and their wider implications in \S \ref{sec: Discussion}.

\section{Equations of motion}
\label{sec: Equations of motion}
\begin{figure}
    \centering
    \includegraphics[width=0.9\textwidth]{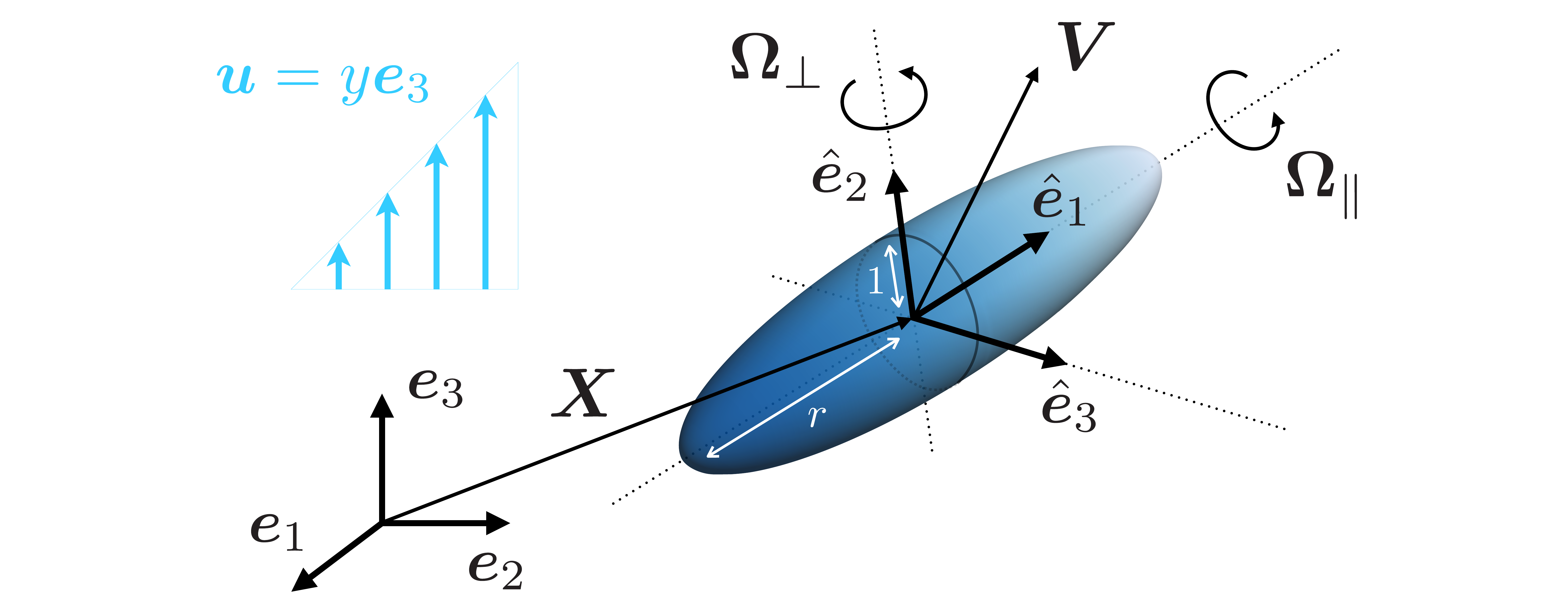}
    \vspace{-2mm}
    \caption{A schematic of the notation and the physical setup we consider in Part 1. We investigate the dynamics of a spheroidal swimmer with equatorial radius $1$, and distance from centre to pole of $\ratio$ along the spheroid symmetry axis $\ehat{1}$. The swimmer has self-generated translational and rotational velocities $\vel = \velscala\ehat{1} + \velscalb\ehat{2} + \velscalc\ehat{3}$ and $\angvel = \omb\ehat{1} + \oma\ehat{2}$, respectively, and interacts with a background shear flow $\vec{u} = y \e{3}$.}
    \label{fig: setup}
\end{figure}

We consider the motion of a rigid, self-propelled spheroid in Stokes flow with imposed far-field shear flow, as illustrated in Figure \ref{fig: setup}. Henceforth, we nondimensionalise time with the inverse shear rate and space with the equatorial radius of the spheroid, so all subsequent variables and parameters are dimensionless with respect to these scalings. The spheroid is therefore defined by a dimensionless equatorial radius of $1$ and distance from centre to pole of $\ratio$ along the spheroid symmetry axis.  In a quiescent fluid, the self-generated swimming velocity of the spheroid is $\vel$, and its self-generated angular velocity is $\angvel$. We assume that both $\vel$ and $\angvel$ are constant in a \emph{swimmer-fixed} reference frame, and that their generating forces and torques are unchanged by the imposition of the external flow. The swimmer-fixed frame will depend on the orientation of the object relative to the laboratory frame, and this orientation depends on $\angvel$. We define these two frames below.

We define the spheroidal axis of symmetry via a swimmer-fixed axis $\ehat{1}$. Then, without loss of generality, we choose $\ehat{2}$ such that $\angvel$ is in a plane spanned by $\ehat{1}$ and $\ehat{2}$, where $\angvel$ makes an angle of $\angl\in[-\pi/2,\pi/2]$ with $\ehat{1}$. We therefore write $\angvel = \omb\ehat{1} + \oma\ehat{2}$, with $\omb$ and $\oma$ being the components of angular velocity that are parallel and perpendicular to the symmetry axis, respectively, and hence $\tan \angl = \oma/\omb$. Then $\ehat{3} = \ehat{1} \times \ehat{2}$.\footnote{If $\angvel \parallel \ehat{1}$, then $\ehat{2}$ and $\ehat{3}$ can be chosen arbitrarily to complete the orthonormal triad.} In this swimmer basis, we write $\vel = \velscala\ehat{1} + \velscalb\ehat{2} + \velscalc\ehat{3}$, while the centre of the spheroid is given by $\Xvecpos(\tl) = \Xpos(\tl)  \e{1} + \Ypos(\tl)  \e{2} + \Zpos(\tl) \e{3}$ with respect to the orthonormal basis $\{\e{1},\e{2},\e{3}\}$ of the laboratory frame. We illustrate these vectors in Figure \ref{fig: setup}.

Finally, we impose a far-field shear flow with velocity field $\flowvel(x,y,z) = y \e{3}$, where $x,y,z$ are the spatial coordinates in the laboratory frame. We are specifically interested in the resultant motion of the swimmer in the presence of this shear flow.

The dynamics for the orientation of the swimmer frame are given in terms of the Euler angles $(\theta, \psi, \phi)$, which we define formally in Appendix \ref{sec: deriving eqs of motion}. Intuitively, one can consider $\theta \in [0, \pi]$ to be the pitch, $\psi \mod{2 \pi}$ to be the roll, and $\phi \mod{2 \pi}$ to be the yaw. By combining the classic equations of \citet{Jeffery1922} with the appropriate rotation axis in terms of Euler angles, the orientation dynamics are governed by
\begin{subequations}
\label{eq: full gov eq}
\begin{align}
\label{eq: theta eq}
\dbyd{\theta}{\tstandard} &= \oma \cos \psi + \fb(\theta,\phi; \Breth), \\
\label{eq: psi eq}
\dbyd{\psi}{\tstandard} &= \omb - \oma \dfrac{\cos \theta \sin \psi}{\sin \theta} + \fc(\theta,\phi; \Breth), \\
\label{eq: phi eq}
\dbyd{\phi}{\tstandard} &= \oma \dfrac{\sin \psi}{\sin \theta} + \fa(\phi; \Breth).
\end{align}
\end{subequations}
where the functions $f_i$ encode the hydrodynamic interaction effects of the flow on the swimmer. For a spheroidal swimmer, \citet{Jeffery1922} showed that these functions are
\begin{subequations}
\label{eq: f functions}
\begin{align}
\fb(\theta,\phi; \Breth) &\coloneqq -\dfrac{\Breth}{2} \cos \theta \sin \theta \sin 2 \phi, \\
\fc(\theta,\phi; \Breth) &\coloneqq \dfrac{ \Breth}{2} \cos \theta \cos 2 \phi, \\
\fa(\phi; \Breth) &\coloneqq \dfrac{1}{2} \left(1 - \Breth \cos 2 \phi\right),
\end{align}
\end{subequations}
where $\Breth = (\ratio^2 - 1)/(\ratio^2 + 1)$ is a constant defined in terms of the spheroidal aspect ratio $\ratio$, and commonly referred to as the Bretherton parameter \citep{Bretherton1962}. For brevity, we will suppress the explicit dependence of the $f_i$ on $\Breth$ unless specifically important. For the chiral swimmers we consider in Part 2, the functions \eqref{eq: f functions} must be modified appropriately to account for the additional generality in swimmer shape. 

Given the symmetries of a spheroidal swimmer, the governing equation for the position of the swimmer in the laboratory frame, $\Xvecpos = \Xpos \e{1} + \Ypos \e{2} + \Zpos \e{3}$, is
\begin{equation}
\label{eq: full gov eq translational}
\dbyd{\Xvecpos}{\tstandard} = \vel + \Ypos \e{3}.
\end{equation}
We emphasise that the simple form of \eqref{eq: full gov eq translational} conceals the full coupling of the system; $\vel = \velscala\ehat{1} + \velscalb\ehat{2} + \velscalc\ehat{3}$ is only constant in the swimmer frame, so the system \eqref{eq: full gov eq translational} is coupled to the angular dynamics through the dependence of the swimmer basis on the Euler angles, which evolve via \eqref{eq: full gov eq}. We also note that the intrinsic symmetries of the spheroid significantly simplify the form of \eqref{eq: full gov eq translational} so that it does not include any contributions involving the rate of strain. We derive and account for these contributions in Part 2, where they are important for the more complex swimmer shapes we investigate therein. As one might expect, those equations for more complex shapes reduce to \eqref{eq: full gov eq}--\eqref{eq: full gov eq translational} in the appropriate limits.

To gain some intuition for the full system \eqref{eq: full gov eq}--\eqref{eq: full gov eq translational}, it is helpful to briefly consider two limiting cases. The first is the absence of rotational propulsion, i.e. $\angvel=\vec{0}$. In this limit,  \eqref{eq: full gov eq}--\eqref{eq: full gov eq translational} reduce to the Jeffery's equations with translation, so the spheroid follows a classic Jeffery's orbit. The other limit is a rotating swimmer (i.e. $\angvel\neq\vec{0}$) in quiescent flow (i.e. $f_i = 0$). Then, invoking the \emph{helix theorem} derived (non-constructively) in \cite{shapere1989geometry} for a general swimmer moving periodically in a quiescent flow, the body will exactly follow a helical path in the laboratory frame. The analytic results we derive in this paper allow us to derive this result constructively in Appendix \ref{sec: helix} for the rotating swimmer we consider. We provide explicit expressions for the radius and pitch of the helical trajectory (dependent on $\angvel$ and $\vel$) that may be of interest for modelling purposes. The goal of this study is to describe the coupled dynamics arising from the complex interaction between the effects of rotation and flow.

Specifically, we systematically investigate how the well-known Jeffery's orbits (which occur when $\angvel=\vec{0}$) should be modified to account for the emergent dynamics of a rapidly rotating active swimmer, which would trace out a helical trajectory in the absence of external flow (which occurs when $f_i = 0$). We are interested in understanding this interaction in the limit where the shear rate of the externally imposed flow is much smaller than the rotation rate of the swimmer, corresponding to $|\angvel| = |\omb\ehat{1} + \oma\ehat{2}| \gg 1$. Additionally, for the translational dynamics, we are interested in the limit where the timescale of translation over the swimmer body radius is much larger than the rotation timescale of the swimmer, corresponding to $|\angvel| \gg |\vel|$. These regimes are relevant to many biological and artificial swimmers, specifically those that achieve propulsion through rapid spinning, as noted in \S \ref{sec: Introduction} for specific examples.

To investigate this interaction, we examine the dynamics that emerge from the nonlinear, autonomous dynamical system \eqref{eq: full gov eq}--\eqref{eq: full gov eq translational}. We treat $\omb > 0$ without loss of generality, allowing $\oma$ to take any real value. The rapid nature of the spinning implies that either $\omb$ or $|\oma|$ (typically both) must be large. In the main analysis of this paper (\S \ref{sec: General emergent behaviour}), we consider rapid generic spinning i.e. $\omb$, $|\oma| \gg 1$ with all other parameters of $\order{1}$. Specifically, we treat $\oma = \order{\omb}$ with $\omb \gg 1$ (which will give the same information as taking $\abs{\angvel} \gg 1$ with $\angl = \order{1}$), which is a \emph{distinguished} asymptotic limit of the system. That is, a general case from which regular asymptotic \emph{sublimits} can be distilled.

However, since the general analysis is fairly technical, requiring a modification of standard multiple scales analysis, we first consider the sublimit of $\omb \gg 1$ and $\oma = \order{1}$ in \S \ref{sec: Bacterial sublimit} (equivalent to $\abs{\angl} \ll 1$). In contrast to the general case, the multiple scales analysis required for this sublimit is standard, and will serve to demonstrate both our notation and the standard methodology. Additionally, this case of rapid axial rotation aligns with a regime common in bacterial swimming, as many bacteria achieve locomotion through the rapid axial rotation of helical flagellar filaments. Readers familiar with the standard method of multiple scales may wish to skip \S \ref{sec: Bacterial sublimit} and proceed directly to \S \ref{sec: General emergent behaviour}, which captures the full range of behaviours of rapidly spinning swimmers via a more general technical analysis. Additionally, readers more interested in the physical implications of our analysis may wish to skip directly to \S \ref{sec: results}, where we summarise and discuss our results in terms of their physical relevance.

\section{Emergent behaviour in the bacterial sublimit}
\label{sec: Bacterial sublimit}

In this section we consider the sublimit of fast bacterial spinning, corresponding to the system \eqref{eq: full gov eq}--\eqref{eq: full gov eq translational} for $\omb \gg 1$ and $\oma = \order{1}$ (equivalent to $\abs{\angl} \ll 1$), with all other parameters of $\order{1}$ (justified on physical grounds). We animate the full dynamics of this system in Supplementary Movie 1. We analyse this system using the standard method of multiple scales, exploiting large $\omb$. As noted above, this sublimit is mathematically straightforward but introduces notation that will be helpful for the more technical general analysis in \S \ref{sec: General emergent behaviour}.

\subsection{Angular dynamics}\label{sec: bacterial sublimit: angular dynamics}
The translational dynamics \eqref{eq: full gov eq translational} decouple from the rest of the system. Hence, we first investigate the angular dynamics given by the system \eqref{eq: full gov eq}--\eqref{eq: f functions} in the $\omb \gg 1$ and $\oma = \order{1}$ limit. To this end, we introduce the fast timescale $\ts$ via
\begin{align}
\label{eq: new times bacterial}
\ts = \omb \tstandard,
\end{align}
and refer to the original timescale $\tstandard$ as the `slow' timescale. Proceeding via the standard method of multiple scales \citep{hinch_1991,Bender1999}, we treat each dependent variable as a function of both the fast and slow timescales, converting ordinary differential equations (ODEs) into partial differential equations (PDEs). The additional degrees of freedom this introduces will be removed later by imposing appropriate periodicity constraints in the fast timescale.

With the new timescale \eqref{eq: new times bacterial}, the time derivative becomes
\begin{align}
\label{eq: time deriv transform bacterial}
\dbyd{}{\tstandard} \mapsto \omb \pbyp{}{\ts} + \pbyp{}{\tl},
\end{align}
which transforms the system \eqref{eq: full gov eq} into
\begin{subequations}
\label{eq: full gov eq trans General bacterial}
\begin{align}
\label{eq: theta eq trans General bacterial}
\omb \pbyp{\theta}{\ts} + \pbyp{\theta}{\tl} &= \oma \cos \psi + \fb(\theta,\phi), \\
\label{eq: psi eq trans General bacterial}
\omb \pbyp{\psi}{\ts} + \pbyp{\psi}{\tl}  &=  \omb - \oma \dfrac{\cos \theta \sin \psi}{\sin \theta} + \fc(\theta,\phi), \\
\label{eq: phi eq trans General bacterial}
\omb  \pbyp{\phi}{\ts} + \pbyp{\phi}{\tl} &= \oma \dfrac{\sin \psi}{\sin \theta} + \fa(\phi).
\end{align}
\end{subequations}

We expand each dependent variable as an asymptotic series in inverse powers of $\omb$, as follows
\begin{align}
\label{eq: asy exp bacterial}
\zeta(\ts,\tl) \sim \zeta_0(\ts,\tl) + \dfrac{1}{\omb} \zeta_1(\ts,\tl) \quad \text{as } \omb \to \infty, \quad \text{ for } \zeta \in \{\theta, \psi, \phi \}.
\end{align}

Using the asymptotic expansions \eqref{eq: asy exp bacterial} in the transformed governing equations \eqref{eq: full gov eq trans General bacterial}, we obtain the leading-order (i.e. $\order{\omb}$) system
\begin{align}
\label{eq: full gov eq trans LO General bacterial}
 \pbyp{\theta_0}{\ts}  = 0, \quad
 \pbyp{\psi_0}{\ts}  = 1, \quad
 \pbyp{\phi_0}{\ts} = 0.
\end{align}
The system \eqref{eq: full gov eq trans LO General bacterial} is straightforward to directly integrate. This will not be the case for the more general problem in \S \ref{sec: General emergent behaviour}, when we have to solve a nontrivial nonlinear problem at leading order. Directly integrating \eqref{eq: full gov eq trans LO General bacterial}, we obtain the solutions
\begin{align}
\label{eq: ang vel BSL}
\theta_0 = \alp(\tl), \quad \psi_0 = \ts + \muc(\tl), \quad \phi_0 = \phic(\tl).
\end{align}
Since $\psi$ and $\phi$ are angles interpreted modulo $2 \pi$, the leading-order solutions \eqref{eq: ang vel BSL} are $2\pi$-periodic in the fast time $\ts$. Importantly, $\alp$, $\muc$, and $\phic$ are as-of-yet undetermined functions of the slow time, directly related to $\theta_0$, $\psi_0$, and $\phi_0$, respectively. Our remaining goal is to determine the governing equations for these slow-time functions. To do this, we must proceed to the next asymptotic order and determine the solvability conditions.

After posing the asymptotic expansions \eqref{eq: asy exp bacterial} and moving the slow-time derivatives to the right-hand side, the $\order{1}$ terms in \eqref{eq: full gov eq trans General bacterial} are
\begin{subequations}
\label{eq: full gov eq trans oma O1 eps bacterial}
\begin{align}
\label{eq: theta eq trans oma O1 eps bacterial}
 \pbyp{\theta_1}{\ts}  &= \oma \cos \psi_0 +  \fb(\theta_0,\phi_0) - \dbyd{\theta_0}{\tl} = \oma \cos (\ts + \muc) + \fb(\alp,\phic) - \dbyd{\alp}{\tl}, \\
 \pbyp{\psi_1}{\ts} &= - \oma \dfrac{\cos \theta_0 \sin \psi_0}{\sin \theta_0} + \fc(\theta_0,\phi_0) - \pbyp{\psi_0}{\tl} \notag \\
 \label{eq: psi eq trans oma O1 eps bacterial}
 &= - \oma \dfrac{\cos \alp \sin (\ts + \muc)}{\sin \alp} + \fc(\alp,\phic) - \dbyd{\muc}{\tl}, \\
 \label{eq: phi eq trans oma O1 eps bacterial}
 \pbyp{\phi_1}{\ts} &= \oma  \dfrac{\sin \psi_0}{\sin \theta_0} + \fa(\phi_0) - \dbyd{\phi_0}{\tl}  = \oma \dfrac{\sin (\ts + \muc)}{\sin \alp} +  \fa(\phic) - \dbyd{\phic}{\tl},
\end{align}
\end{subequations}
now accompanied by fast-time periodicity constraints over any $2\pi$ period in $\ts$. These remove the additional degrees of freedom introduced by initially taking $\ts$ and $\tl$ to be independent. For the system \eqref{eq: full gov eq trans oma O1 eps bacterial}, it is straightforward to determine the requisite solvability conditions. This is because the linear operators acting on $\theta_1$, $\psi_1$, and $\phi_1$ (on the left-hand sides) decouple from one another, even though the full problem \eqref{eq: full gov eq} is fully coupled. This will not be the case for the more general problem in \S \ref{sec: General emergent behaviour}, when we have to derive and solve a nontrivial adjoint problem.

Nevertheless, for the sublimit of bacterial spinning we consider in this section, this decoupling means that we can straightforwardly obtain our desired solvability conditions by integrating \eqref{eq: full gov eq trans oma O1 eps bacterial} with respect to $\ts$ from $0$ to $2 \pi$. The integrated fast-time derivatives on the left-hand side vanish due to the imposed periodicity, and the terms multiplied by $\oma$ on the right-hand sides also vanish due to their specific trigonometric forms. Therefore, this integration yields the system
\begin{align}
\label{eq: full gov eq trans oma O1 eps bacterial int}
\dbyd{\alp}{\tl} = \fb(\alp,\phic;\Breth), \quad
\dbyd{\muc}{\tl} = \fc(\alp,\phic;\Breth), \quad
\dbyd{\phic}{\tl} = \fa(\phic;\Breth),
\end{align}
which provide the governing equations we seek for the slow-time functions in the leading-order solutions \eqref{eq: ang vel BSL}. Importantly, we see that the slow-time system \eqref{eq: full gov eq trans oma O1 eps bacterial int} is equivalent to the full system \eqref{eq: full gov eq} without the rotation terms (i.e. $\omb = \oma = 0$) using the equivalence $(\theta, \psi, \phi) \leftrightarrow (\alp, \muc, \phic)$. One consequence of this is that $\muc$ decouples from $\alp$ and $\phic$, since $\psi$ decouples from $\phi$ and $\theta$ in the absence of rotation. This property is maintained in the more general analysis we present in \S \ref{sec: General emergent behaviour}. Moreover, we note that the Bretherton parameter $\Breth$ remains unchanged in the emergent behaviour in this limit. This property is not maintained in \S \ref{sec: General emergent behaviour}.

To dwell on the general point for clarity, in the sublimit of fast bacterial spinning (with $\omb \gg 1$ and $\oma = \order{1}$) the emergent behaviour is the same as it would be without any fast spinning.  While $\psi$ does vary rapidly due to this fast spinning, the leading-order emergent angular dynamics (up to and including timescales of $\tstandard = \order{1}$) are not otherwise affected by the fast spinning. Importantly, this emergent indifference to fast spinning is not the case for the more general rotation we consider in \S \ref{sec: General emergent behaviour}.

\begin{figure}
    \centering
    \vspace{2em}
    \begin{overpic}[width=0.95\textwidth]{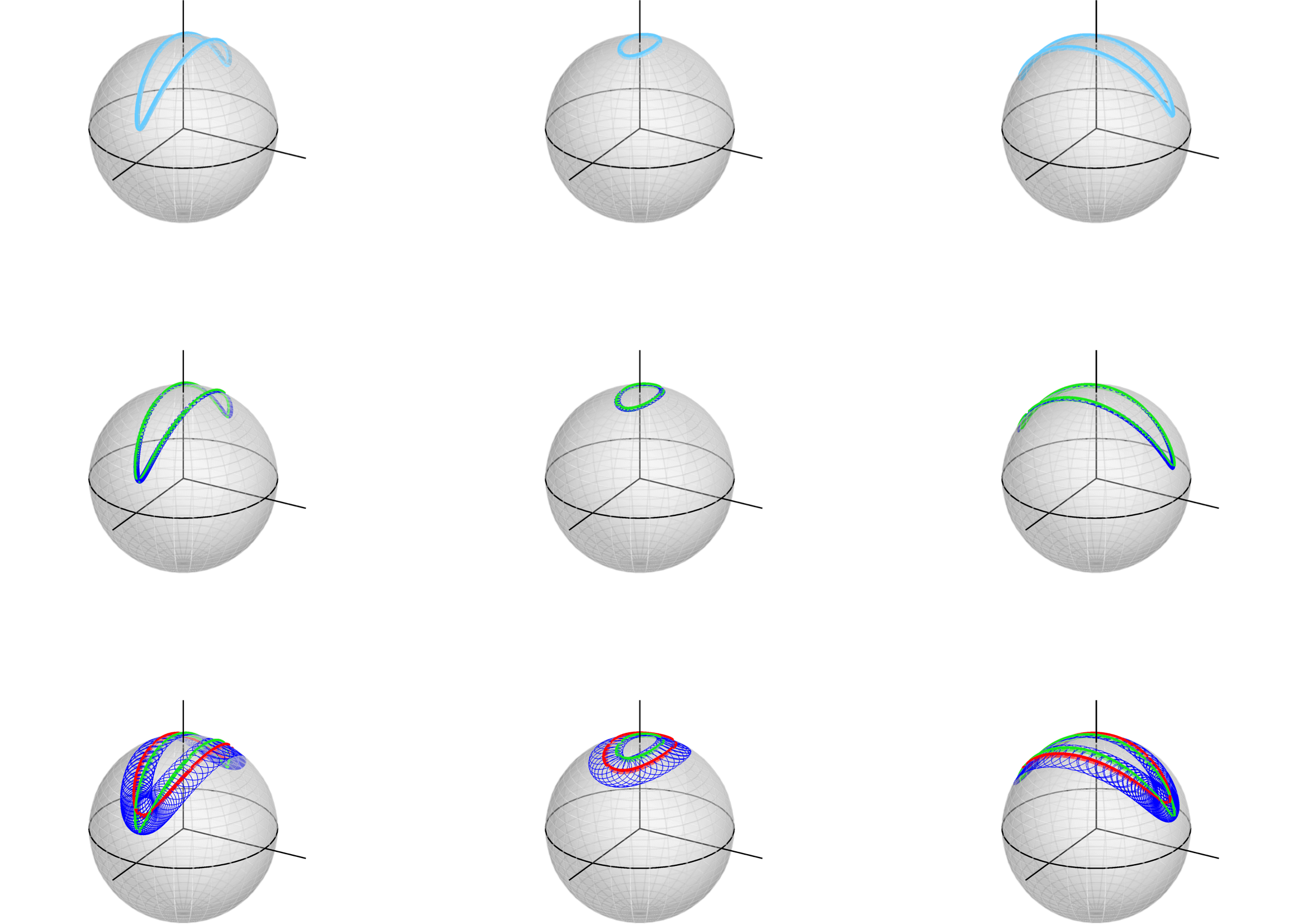}
    \put(38,72){(a) $\omb = \oma = 0$}
    \put(9,51){$\Breth = 0.99$}
    \put(45,51){$\Breth = 0.5$}
    \put(78,51){$\Breth = -0.99$}
    \put(36,46){(b) $\omb = 15$, $\oma = 0.5$}
    \put(9,24){$\Breth = 0.99$}
    \put(45,24){$\Breth = 0.5$}
    \put(78,24){$\Breth = -0.99$}
    \put(37,19){(c) $\omb = 15$, $\oma = 3$}
    \put(9,-3){$\Breth = 0.99$}
    \put(45,-3){$\Breth = 0.5$}
    \put(78,-3){$\Breth = -0.99$}
    \end{overpic}
    \vspace{2em}
    \caption{Rotational dynamics in the bacterial limit. Each plot displays the evolution of the particle orientation on the unit sphere, parameterised with $(\theta,\phi)$ as standard spherical coordinates, with the vertical axis corresponding to $\theta=0$. (a) Standard 3D Jeffery's orbits of passive particles for three different values of the Bretherton parameter $\Breth = (\ratio^2 - 1)/(\ratio^2 + 1)$. (b) Incorporating active spinning in the bacterial limit, with $\omb \gg 1, \oma = \order{1}$. The full particle orientation dynamics are shown as a thin blue line, while the thicker green line shows the solution for the emergent, averaged behaviour calculated in \S\ref{sec: Bacterial sublimit}. Notably, this thick green line traces the same orbit as the blue curves in (a); as predicted by \eqref{eq: full gov eq trans oma O1 eps bacterial int}, the leading-order dynamics are not affected by spinning in this limit. (c) Further increasing $\oma$ beyond the bacterial limit causes a loss in the validity of our predictions thus far. The full evolution of the particle orientation \eqref{eq: full gov eq}--\eqref{eq: f functions} is shown by the thin blue line, while the emergent dynamics predicted by \eqref{eq: full gov eq trans oma O1 eps bacterial int} are shown by the thick green line, in poor agreement with the full dynamics. Accurately capturing these emergent trajectories requires the more general analysis of \S\ref{sec: General emergent behaviour}, the predictions of which are shown as red curves.}
    \label{fig:rotational_BL}
\end{figure}
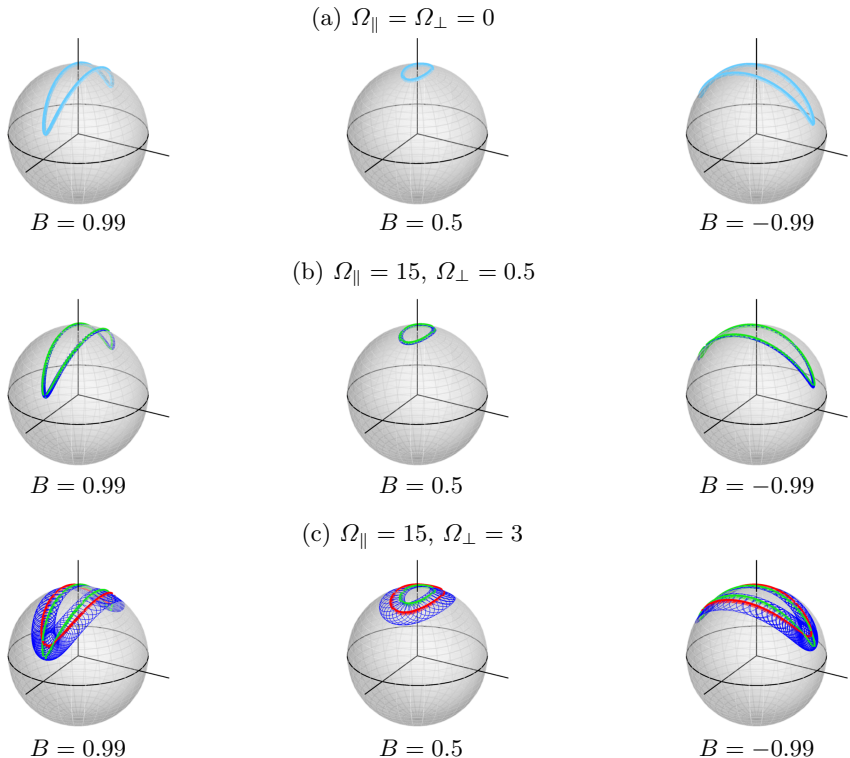

\Cref{fig:rotational_BL} illustrates these observations through numerical solution of the full and the systematically averaged governing equations, displaying the evolution of the swimmer's orientation in several different cases. In this figure, we illustrate the changing orientation of the object by plotting a trajectory on the surface of the unit sphere, which $\theta$ and $\phi$ naturally parameterise via standard spherical coordinates. With the rapid evolution of $\psi$ not illustrated, the particle's orientation traces out a closed orbit on the sphere, corresponding to a Jeffery's orbit that depends on the Bretherton parameter $\Breth$ and the initial conditions. The orbits of passive particles, with $\oma=\omb=0$, are illustrated in (a), with varying shape parameters and initial conditions ($\theta(0)=\pi/18$, $\psi(0)=-\pi/4$ in all columns, and $\phi(0)=\pi/2$ in all but the final column, where $\phi(0) = 0$). We see from (a) and (b) that the emergent angular dynamics are not affected by the fast spinning as long as $\oma$ is not significant, and so active spinning particles generate the same orbits as passive particles \emph{if} the off-axis spinning is not significant. However, we see from (c) that this result breaks down once $\oma$ starts to become a little larger; for an active particle with a significant component of off-axis spinning, the evolution of $\phi$ and $\psi$ differs from that of a passive particle. 

In more detail, in \Cref{fig:rotational_BL}b the swimmer spins with $\omb=15$ and $\oma=0.5$ and its orientation over time is represented as a thin dark blue line. The trajectory of the leading-order dynamics is shown as a thick green line, and can be seen to be in excellent agreement with the average evolution of the swimmer's orientation; moreover, it is indeed essentially identical to the non-spinning Jeffery orbit from \Cref{fig:rotational_BL}a, in line with the conclusions of our asymptotic analysis.

The analysis in this section starts to lose validity when the magnitude of the off-axis rotation $\oma$ increases, since this voids the requirement $\oma = \order{1}$. This can be seen in \Cref{fig:rotational_BL}c, where the dynamics predicted by \eqref{eq: full gov eq trans oma O1 eps bacterial int} start to diverge from the actual emergent behaviour of the swimmer, here with $\omb=15$ and $\oma=3$. As before, the blue line shows the full dynamics of the swimmer, and the green line shows the emergent dynamics predicted by the analysis of this section. The latter fails to well-approximate the average evolution of the orientation since $\oma$ is now significant. In \S \ref{sec: General emergent behaviour}, we generalise our analysis to capture the emergent behaviour outside the bacterial sublimit considered in this section. The predictions of this later analysis are shown as red curves in \Cref{fig:rotational_BL}c, which exhibit excellent agreement with the averaged behaviour of the full system.

We have now calculated the emergent behaviour of the angular dynamics in the sublimit of bacterial spinning. We close this problem by considering the emergent behaviour of the translational dynamics.

\subsection{Translational dynamics}\label{sec: bacterial sublimit: translational dynamics}
The governing equations for translation are given in \eqref{eq: full gov eq translational}. Using the transformation \eqref{eq: time deriv transform bacterial}, the translational governing equations \eqref{eq: full gov eq translational} become
\begin{multline}
\label{eq: transform gov eq translational BL}
    \omb \pbyp{\Xvecpos}{\ts} + \pbyp{\Xvecpos}{\tl} =  \velscala\ehat{1}(\theta, \phi) + \velscalb\ehat{2}(\theta,\psi, \phi) + \velscalc\ehat{3}(\theta,\psi, \phi) + \Ypos \e{3},
\end{multline}
with the $\ehat{i}$ denoting the swimmer frame basis vectors, given in terms of the laboratory frame basis vectors $\e{i}$ in \eqref{elm2}.

Following \eqref{eq: asy exp bacterial} and expanding each dependent variable as an asymptotic series in inverse powers of $\omb$, we obtain the $\mathit{O}(\omb)$ (leading-order) system
\begin{align}
\label{eq: transform gov eq translational LO BL}
\pbyp{\Xvecpos_0}{\ts} = \boldsymbol{0}.
\end{align}
Hence, the vector position $\Xvecpos$ is independent of the fast time $\ts$ at leading order. That is, $\Xvecpos_0 = \Xvecpos_0(\tl)$. Our remaining goal is to derive the dependence of $\Xvecpos_0$ on the slow time via the calculation of solvability conditions at higher order.

At next order (i.e. $\mathit{O}(1)$), the translational governing equations \eqref{eq: transform gov eq translational BL} are
\begin{align} 
\label{eq: transform gov eq translational FC BL}
    \pbyp{\Xvecpos_1}{\ts} + \dbyd{\Xvecpos_0}{\tl} = \velscala \ehat{1}(\theta_0,\phi_0)  + \velscalb \ehat{2}(\theta_0,\psi_0,\phi_0) + \velscalc \ehat{3} (\theta_0,\psi_0,\phi_0) + \Ypos_0 \e{3},
\end{align}
along with fast-time periodicity constraints over any $2 \pi$ period in $\ts$. It is straightforward to obtain the appropriate solvability conditions by integrating \eqref{eq: transform gov eq translational FC BL} with respect to the fast time $\ts$ from $0$ to $2 \pi$. In the same way as in \S \ref{sec: bacterial sublimit: angular dynamics}, the fast-time derivative on the left-hand side vanishes due to the imposition of fast-time periodicity. Substituting the leading-order angular dynamics solutions \eqref{eq: ang vel BSL} into the explicit forms of $\ehat{i}$ given in \eqref{elm2}, we find that
\begin{align}
\label{eq: av basis vectors in BL}
    \av{\ehat{1}(\theta_0,\phi_0)} = \etilde{1}(\alp,\phic), \quad
    \av{\ehat{2}(\theta_0,\psi_0,\phi_0)} = \vec{0}, \quad
    \av{\ehat{3}(\theta_0,\psi_0,\phi_0)} = \vec{0},
\end{align}
using the notation $\av{\bcdot}$, defined as the average of its argument over one fast-time oscillation
\begin{equation}
\label{eq: av operator}
    \av{y} = \dfrac{1}{2 \pi}\int_0^{2\pi} \! y \,  \mathrm{d}\ts.
\end{equation}
Additionally, we note that $\etilde{1}(\alp,\phic)$ can be considered equivalent to the (hatted) basis vector $\ehat{1}$ in \eqref{elm2}, but with argument $(\theta,\phi)$ replaced by $(\alp,\phic)$.

Hence, using the results \eqref{eq: av basis vectors in BL}, taking the fast-time average of \eqref{eq: transform gov eq translational FC BL} yields
\begin{equation}
\label{eq: transform gov eq translational emergent0}
    \dbyd{\Xvecpos_0}{\tl} = \velscala \etilde{1}(\alp,\phic) + \Ypos_0 \e{3}.
\end{equation}
In particular, comparing the slow-time equation \eqref{eq: transform gov eq translational emergent0} to the full translational dynamics of \eqref{eq: full gov eq translational} highlights that the structure of the translational dynamics is broadly unchanged by the rapid spinning, with basis vectors being replaced by their averages. However, this means that the self-induced velocity components in the original $\ehat{2}(\theta,\psi,\phi)$ and $\ehat{3}(\theta,\psi,\phi)$ directions are cancelled out by the fast spinning of the swimmer, so that only the velocity component in the original $\ehat{1}(\theta,\phi)$ direction contributes to the average velocity alongside the shear flow. Hence, this analysis shows that one can effectively neglect the off-axis components of the self-induced propulsive velocity of such a swimmer, at least at leading order and for $\tstandard = \order{1}$. Interpreted physically, this result establishes that rapid axial spinning, in the absence of significant off-axis rotation, is sufficient to justify modelling the translation of such a swimmer using only the axial component of propulsive linear velocity, since the other self-velocity contributions cancel out at leading order. We show in \S \ref{etd} that this intuitive result requires modification when off-axis rotation is significant.

\begin{figure}
    \centering
    \vspace{1em}
    \begin{overpic}[width=\textwidth]{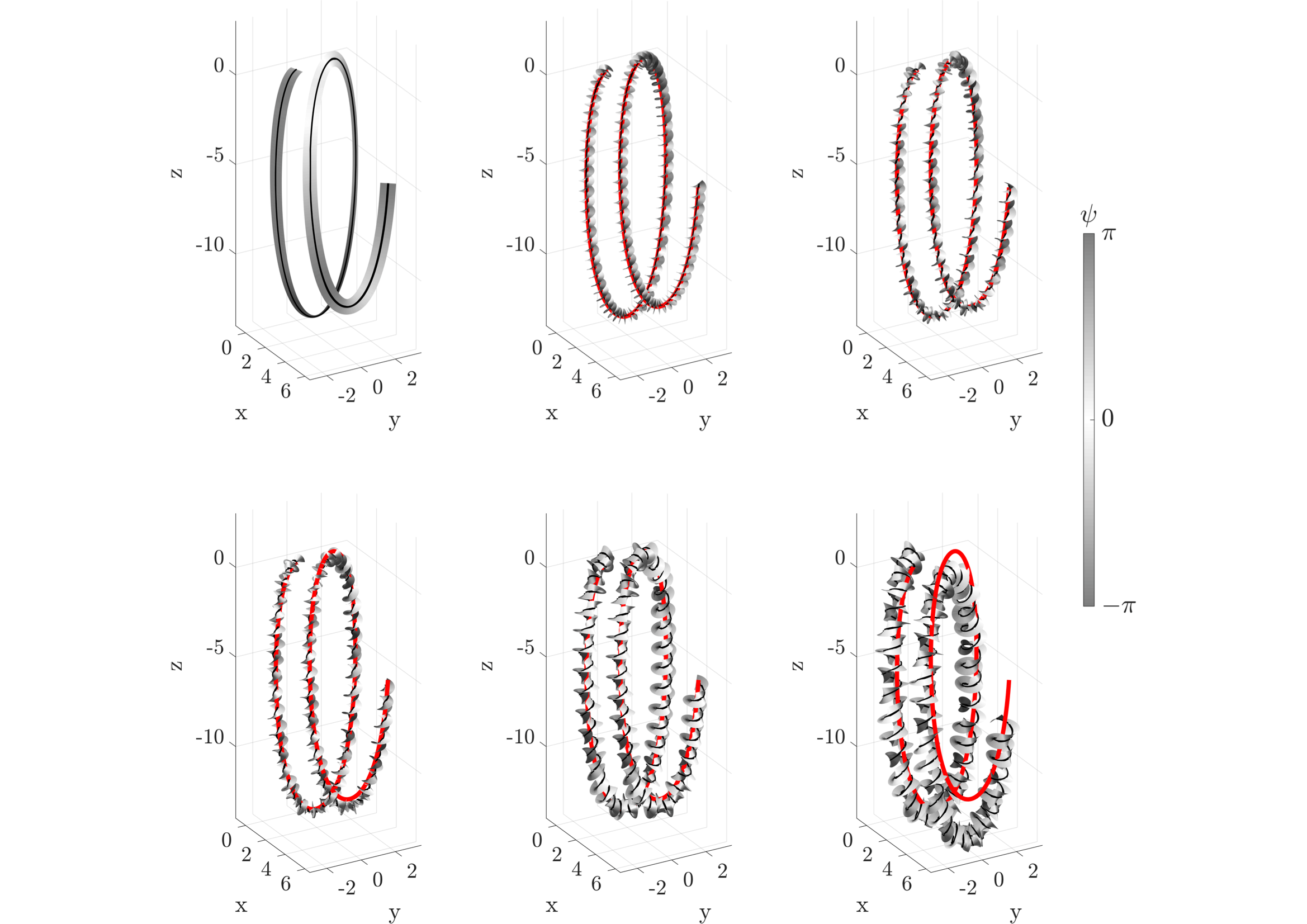}
    \put(12,70){(a)}
    \put(37,70){(b)}
    \put(60,70){(c)}
    \put(12,30){(d)}
    \put(37,30){(e)}
    \put(60,30){(f)}
    \end{overpic}
    \caption{
    Translational dynamics of a spheroidal swimmer in the bacterial limit. A dynamic version of this figure is given in Supplementary Movie 1. In each panel, the black line traces the swimmer position $\Xvecpos$ over time via \eqref{eq: full gov eq}--\eqref{eq: full gov eq translational}, with $\Xvecpos = \vec{0}$ and $(\theta,\psi,\phi) = (2\pi/5,\pi/2,-7\pi/15)$ initially. Here, we have fixed $\Breth=0.5$ and $\velscala=1$. The shaded ribbons represent the evolution of the swimmer orientation, with its twist and shading encoding the value of $\psi$ according to the colourbar. In panel (a), the body does not have an intrinsic rotation ($\omb = \oma = 0$), nor any off-axis propulsive velocities ($\velscalb = \velscalc = 0$). In all other panels, the particle is spinning with $\omb = 15$ (and $\oma = 0$) and the thick red line shows the predicted translational dynamics \eqref{eq: full gov eq trans oma O1 eps bacterial int}, \eqref{eq: transform gov eq translational emergent0}, in excellent agreement with full dynamics. The panels differ in their prescribed off-axis propulsive velocities: (b) $\velscalb = \velscalc = 0$; (c) $\velscalb = 1, \velscalc = 0$; (d) $\velscalb = \velscalc = 1$; (e) $\velscalb = 5, \velscalc = 0$; (f) $\velscalb = \velscalc = 5$. Despite the range of prescribed propulsive velocities and correspondingly intricate fast-timescale trajectories seen here, the emergent average translation is consistent between panels, as predicted by our analysis.}
    \label{fig:translational_BL}
\end{figure}
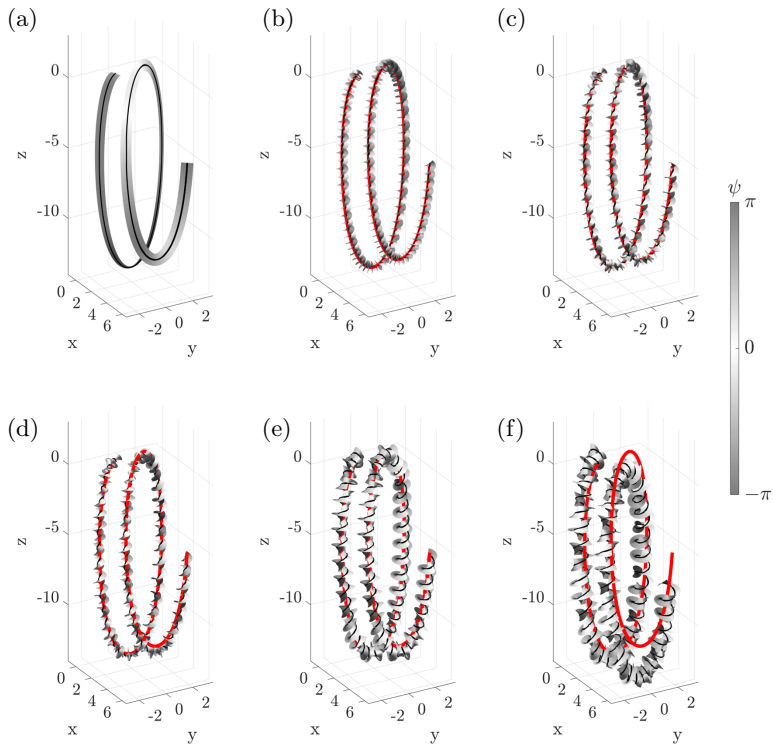

We show some examples of the swimming behaviours of actively spinning particles in \Cref{fig:translational_BL} and Supplementary Movie 1. Each panel displays the evolution of a swimmer's position as a black line. A superposed ribbon illustrates the evolution of the intrinsic orientation angle $\psi$; its shade of grey indicates the value of $\psi \in [-\pi, \pi]$ according to the colourbar on the right-hand side of the figure.

\Cref{fig:translational_BL}a showcases the relatively simple dynamics of a particle that is not actively spinning ($\omb=\oma=0$) and is only propulsing along its symmetry axis ($\velscala = 1$, $\velscalb = \velscalc = 0$), while \Cref{fig:translational_BL}b--f illustrate the behaviours of spinning particles with various propulsive velocities, having fixed $\velscala = 1$ in each panel and taking $\omb=15$ and $\oma=0$ in all but \Cref{fig:translational_BL}a. In each panel, the trajectories predicted by the asymptotic analysis are shown as thick red lines, showcasing excellent agreement with the full dynamics in all but the final case, as expected, wherein the components of propulsive velocity become comparable in magnitude to the fast rate of active rotation.

The near-indistinguishable nature of the red and black lines in panel (b), where $\velscalb = \velscalc = 0$, highlights that the rapid spinning illustrated by the twisting ribbon need not significantly modify the emergent translational dynamics when there is no off-axis propulsion. In panels (c) to (e) we allow $(\velscalb,\velscalc)\neq(0,0)$, which causes far more complex fast-scale dynamics, as can be seen by the cork-screwing of the black line. However, as predicted by our asymptotic analysis, the combination of active spinning and non-trivial self-propulsion does not materially modify the predicted average emergent dynamics (red line). A slight exception to this is shown in \Cref{fig:translational_BL}f, which corresponds to a case where the propulsive velocity is no longer asymptotically smaller than the rate of the rapid rotation. This starts to break the requirement of a separation of scales, giving rise to a discrepancy between the predictions of the asymptotic analysis and the full dynamics. 
 
\section{Deriving the general emergent behaviour}
\label{sec: General emergent behaviour}

We now consider the more general problem, the distinguished asymptotic limit where $\omb$, $|\oma| \gg 1$ with $\oma = \order{\omb}$ ($\angl = \order{1}$), and all other parameters are of $\order{1}$. We animate the full dynamics of this system in Supplementary Movie 2. As before, we treat $\omb > 0$ without loss of generality, but allow $\oma$ to take any real value. In this distinguished limit, the subcases of $|\oma| \ll \omb$ and $|\oma | \gg \omb$ can be obtained through taking appropriate limits of the results we derive. Given the above, it will be helpful to introduce the notation $\ac = \order{1}$ such that
\begin{align}\label{eq: def omega as ratio of Omegas}
\oma = \ac \omb,
\end{align}
and to formally consider the single asymptotic limit $\omb \gg 1$. We emphasise that $\ac$ can take any real value, and that the angle between the symmetry and rotational axes $\angl$ is related to $\ac$ through $\tan \angl = \ac$. Finally, we treat all other parameters as $\order{1}$, justified on physical grounds.

We therefore analyse the system \eqref{eq: full gov eq}--\eqref{eq: full gov eq translational} using a modified method of multiple scales for systems in the limit of large $\omb$, treating $\oma = \order{\omb}$. Our goal is to derive effective equations for the emergent dynamics, in terms of the system parameters. To this end, we introduce the fast timescale $\ts$ as follows
\begin{align}
\label{eq: new times}
\ts = \left(\oma^2 +\omb^2\right)^{1/2} \tstandard = \om \omb \tstandard,
\end{align}
where we use
\begin{align}
\om \coloneqq \sqrt{1 +\ac^2},
\end{align}
for notational convenience, and we refer to the original timescale $\tstandard$ as the `slow' timescale. Under the transformations \eqref{eq: new times}, the time derivative becomes
\begin{align}
\label{eq: time deriv transform}
\dbyd{}{\tstandard} \mapsto \om \omb \pbyp{}{\ts} + \pbyp{}{\tl}.
\end{align}

We note that the fast-time notation $\ts$ in \eqref{eq: new times} is slightly more general than $\ts$ from the previous section (defined in \eqref{eq: new times bacterial}), but that there is a regular limit in which \eqref{eq: new times} reduces to \eqref{eq: new times bacterial}. Specifically, the bacterial sublimit in the previous section can be obtained formally from the analysis in this section by considering the limit $\ac \to 0$, and hence $\om \to 1$. We proceed as before, treating each dependent parameter as a function of both new timescales. Again, we later remove the additional degrees of freedom this introduces by imposing appropriate periodicity constraints in the fast timescale. 

\subsection{Leading-order angular dynamics}
\label{sec: leading-order}

For our analysis of the general problem, we start by considering the angular dynamics. The time-derivative transformation \eqref{eq: time deriv transform} converts the system \eqref{eq: full gov eq} into
\begin{subequations}
\label{eq: full gov eq trans General}
\begin{align}
\label{eq: theta eq trans General}
 \om \omb \pbyp{\theta}{\ts} + \pbyp{\theta}{\tl} &= \ac \omb \cos \psi + \fb(\theta,\phi), \\
\label{eq: psi eq trans General}
 \om \omb \pbyp{\psi}{\ts} + \pbyp{\psi}{\tl}  &= \omb\left(1 - \ac \dfrac{\cos \theta \sin \psi}{\sin \theta}\right) + \fc(\theta,\phi), \\
 \label{eq: phi eq trans General}
\om \omb \pbyp{\phi}{\ts} +  \pbyp{\phi}{\tl} &= \ac \omb \dfrac{\sin \psi}{\sin \theta} + \fa(\phi).
\end{align}
\end{subequations}

We expand each dependent variable as an asymptotic series in inverse powers of $\omb$, following \eqref{eq: asy exp bacterial}. Then the leading-order (i.e. $\order{\omb}$) version of \eqref{eq: full gov eq trans General} is
\begin{subequations}
\label{eq: full gov eq trans LO General}
\begin{align}
\label{eq: theta eq trans LO General}
\om \pbyp{\theta_0}{\ts}  &= \ac\cos \psi_0, \\
\label{eq: psi eq trans LO General}
\om \pbyp{\psi_0}{\ts}  &= 1 - \ac \dfrac{\cos \theta_0 \sin \psi_0}{\sin \theta_0}, \\
\label{eq: phi eq trans LO General}
\om \pbyp{\phi_0}{\ts} &= \ac\dfrac{\sin \psi_0}{\sin \theta_0}.
\end{align}
\end{subequations}
The leading-order system \eqref{eq: full gov eq trans LO General} is forced solely by the rapid spinning of the swimmer, and constitutes a nonlinear 3D dynamical system for $\bs{x}_0 \coloneqq (\theta_0, \psi_0, \phi_0)^{T}$, which is notation we will use later.

A key difference in our analysis of the general problem in this section is that the nonlinear, coupled leading-order system \eqref{eq: full gov eq trans LO General} is not straightforward to solve. This is in direct contrast to the linear, decoupled leading-order system \eqref{eq: full gov eq trans LO General bacterial} of the previous section. One consequence of this is that the important slow-time variables are not immediately obvious.

Of great help in our task is the fact that we can solve the leading-order system \eqref{eq: full gov eq trans LO General} in closed form. This is abetted by two key observations. Firstly, \eqref{eq: phi eq trans LO General} decouples from the other two subequations, and hence we treat it last. Secondly,  \eqref{eq: theta eq trans LO General} and \eqref{eq: psi eq trans LO General} can be combined and integrated to admit a first integral, which constitutes one of the three slow-time functions we expect to arise. To see this, we first divide \eqref{eq: psi eq trans LO General} by \eqref{eq: theta eq trans LO General} and multiply through by $\ac\cos \psi_0$ to yield
\begin{align}
\label{eq: quotient derivative}
\ac\cos \psi_0  \pbyp{\psi_0}{\theta_0}
= 1 - \ac \dfrac{\cos \theta_0 \sin \psi_0}{\sin \theta_0}.
\end{align}
Since the left-hand side of \eqref{eq: quotient derivative} can be written as $\ac \partial (\sin \psi_0)/\partial \theta_0$, the \cref{eq: quotient derivative} can be rewritten as a first-order linear differential equation for $\sin \psi_0$ as a function of $\theta_0$ i.e. implicitly making the change of variables $(\ts,\tl) \mapsto (\theta_0,\tl)$. Multiplying through by the appropriate integrating factor of $\sin \theta_0$ and grouping everything on the same side, we may re-write \eqref{eq: quotient derivative} as
\begin{align}
\label{eq: H halfway}
0 = \dfrac{\partial}{\partial \theta_0} \left(\ac \sin \theta_0 \sin \psi_0 \right) - \sin \theta_0.
\end{align}
Hence, the required first integral is obtained by directly integrating \eqref{eq: H halfway} to yield
\begin{align}
\label{eq: H with H}
H(\tl) = \ac \sin \theta_0 \sin \psi_0 +  \cos \theta_0,
\end{align}
thus generating the function of integration $H(\tl)$, emphasising that $H$ is independent of the fast time $\ts$. As will become clear once we complete the full analysis, it will turn out to be particularly convenient to redefine $H(\tl) = \om \cos \alp(\tl)$, so we formally write
\begin{align}
\label{eq: alpha def}
 \om \cos \alp(\tl) =\ac \sin \theta_0 \sin \psi_0 +  \cos \theta_0,
\end{align}
where $\alp = \alp(\tl)$ is the first of three slow-time functions (each marked with an overbar) we obtain from our leading-order analysis in this section. We note that our definition of $\alp$ in \eqref{eq: alpha def} in the limit $\ac \to 0$ coincides with its former definition in \eqref{eq: ang vel BSL}. The goal of the next-order analysis in \S \ref{sec: FC} is to derive the governing equations satisfied by $\alp$ and the two other appropriate slow-time functions we derive in the remainder of this section.

Substituting the fast-time-conserved quantity \eqref{eq: alpha def} into \eqref{eq: theta eq trans LO General}, we obtain the following nonlinear first-order differential equation:
\begin{align}
\label{eq: Eq for cos theta 0}
\om \pbyp{\theta_0}{\ts} &=  \dfrac{\pm\sqrt{\ac^2 \sin^2 \theta_0 -  \left( \om \cos \alp - \cos \theta_0 \right)^2}}{\sin \theta_0}.
\end{align}
Since \eqref{eq: Eq for cos theta 0} is an autonomous first-order differential equation for $\theta_0$ as a function of $\ts$, it admits a solution in the implicit form
\begin{align}
\label{eq: mu integral}
\pm \int^{\theta_0} \! \dfrac{\om \sin s  \, \mathrm{d} s}{\sqrt{\ac^2 \sin^2 s -  \left( \om \cos \alp - \cos s \right)^2}} = \ts + \muc(\tl),
\end{align}
where $\muc = \muc(\tl)$ is the second of the three slow-time functions we obtain from our leading-order analysis. In fact, we can evaluate \eqref{eq: mu integral} explicitly using the substitution $\om \cos s = \cos \alp + u \sin \alp$ to  convert \eqref{eq: mu integral} into
\begin{align}
\label{eq: mu integral easier}
\mp \int \! \dfrac{\mathrm{d} u}{\sqrt{\ac^2  - u^2}} = \ts + \muc,
\end{align}
omitting the transformed limit of the integral for notational convenience. It is then straightforward to integrate the left-hand side of \eqref{eq: mu integral easier} and rearrange to deduce that
\begin{subequations}
\label{eq: mu def}
\begin{align}
\label{eq: om cos thet def}
\om \cos \theta_0 = \cos \alp - \ac \sin \alp \cos (\ts + \muc).
\end{align}
We note that the $\mp$ in \eqref{eq: mu integral easier} (as well as the choice of $\cos (\ts  +\muc)$ in \eqref{eq: om cos thet def} instead of $\sin (\ts  +\muc)$) can be absorbed into an initial phase shift in $\muc$. Therefore, we have taken the negative root in moving from \eqref{eq: mu integral easier} to \eqref{eq: om cos thet def} for later convenience, without loss of generality. Rearranging \eqref{eq: om cos thet def}, we deduce that
\begin{align}
\label{eq: om sin thet sq def}
\om^2 \sin^2 \theta_0 = (\ac \cos \alp \cos(\ts + \muc) + \sin \alp)^2 + 
\ac^2 \sin^2 (\ts + \muc).
\end{align}
It will also be helpful later to note that the combination of \eqref{eq: alpha def} and \eqref{eq: om cos thet def} yields the relationship
\begin{align}
\label{eq: sin theta sin psi}
\om \sin \theta_0 \sin \psi_0 = \ac \cos \alp + \sin \alp \cos (\ts + \muc),
\end{align}
and that differentiating \eqref{eq: om cos thet def} with respect to $\ts$ and invoking \eqref{eq: theta eq trans LO General} yields the relationship
\begin{align}
\label{eq: sin theta cos psi}
\sin \theta_0 \cos \psi_0 &= -\sin \alp \sin (\ts + \muc).
\end{align}
\end{subequations}
Given the relationships \eqref{eq: mu def}, we see that the trigonometric quantities $\cos \theta_0$, $\sin \theta_0 \sin \psi_0$, and $\sin \theta_0 \cos \psi_0$ are fast oscillations in time whose mean, amplitude, and phase are slowly varying, coupled quantities.

The final result required from the leading-order system is the solution of \eqref{eq: phi eq trans LO General}, and the subsequent appearance of the third (and final) slow-time function. Re-writing the right-hand side of \eqref{eq: phi eq trans LO General} as $\sin \theta_0 \sin \psi_0/\sin^2 \theta_0$, we can use the expressions \eqref{eq: om sin thet sq def}--\eqref{eq: sin theta sin psi} to re-write \eqref{eq: phi eq trans LO General} as
\begin{align}
\label{eq: phi eq trans LO General explicit}
\pbyp{\phi_0}{\ts} = \dfrac{\ac^2 \cos \alp + \ac\sin \alp \cos (\ts + \muc)}{(\ac \cos \alp \cos (\ts + \muc) + \sin \alp)^2 + \ac^2 \sin^2 (\ts + \muc)}.
\end{align}
Then, observing that
\begin{align}
\dfrac{\partial}{\partial \ts}\left( \dfrac{\ac \sin (\ts + \muc)}{\ac \cos \alp \cos (\ts + \muc) + \sin \alp} \right) = \dfrac{\ac^2 \cos \alp + \ac\sin \alp \cos (\ts + \muc)}{(\ac \cos \alp \cos (\ts + \muc) + \sin \alp)^2},
\end{align}
integrating \eqref{eq: phi eq trans LO General explicit} directly with respect to the variable $\ac \sin (\ts + \muc) / (\ac \cos \alp \cos (\ts + \muc) + \sin \alp)$ yields the solution
\begin{align}
\label{eq: phi0 sol}
\tan(\phi_0 - \phic(\tl)) = \dfrac{\ac \sin (\ts + \muc)}{\ac \cos \alp \cos (\ts + \muc) + \sin \alp},
\end{align}
where $\phic = \phic(\tl)$ is the final slow-time function we obtain from our leading-order analysis. For later convenience, it is also helpful to expand out the left-hand side of \eqref{eq: phi0 sol} and rearrange to obtain the relationship
\begin{align}
\label{eq: tan phi0}
\tan \phi_0 &= \dfrac{\ac \cos \phic \sin (\ts + \muc) + \sin \phic \left(\ac \cos \alp \cos (\ts + \muc) + \sin \alp \right)}{\cos \phic \left(\ac \cos \alp \cos (\ts + \muc) + \sin \alp \right) - \ac \sin \phic \sin (\ts + \muc)}.
\end{align}

We have now fully solved the nonlinear, coupled leading-order problem \eqref{eq: full gov eq trans LO General} to obtain the fast-time solutions \eqref{eq: mu def} and \eqref{eq: tan phi0}, in terms of the three slow-time functions of integration $\alp(\tl)$, $\muc(\tl)$, and $\phic(\tl)$. Importantly for the method of multiple scales, the leading-order solutions given in \eqref{eq: mu def} and \eqref{eq: tan phi0} are $2 \pi$-periodic in the fast time $\ts$ (appropriately interpreting $\psi_0$ and $\phi_0$ modulo $2 \pi$).\footnote{Despite starting with a nonlinear oscillator system, we are able to use multiple scales instead of the more general method of \citet{kuzmak1959asymptotic} because the period of the fast-time oscillation is independent of the slow time.} The slow-time functions $\alp(\tl)$, $\muc(\tl)$, and $\phic(\tl)$ are currently undetermined, and the goal of our next-order analysis in \S \ref{sec: FC} is to derive the governing equations for these functions. For generic initial conditions of the full system, one can also derive the appropriate initial conditions for these slow-time functions. We present this derivation in Appendix~\ref{sec: Slow time functions IC}, noting that some care is required in order to ensure that the correct branches are taken.

Over the slow time, one can think of $\alp$ as controlling some emergent amplitude or mean of oscillation, $\muc$ as controlling some emergent phase of oscillation, and $\phic$ as an emergent drift in yawing. Additionally, in a sense to be made precise later, we can interpret each slow-time function as being associated with an underlying variable. Specifically, and in the same way as their equivalent variables in \S\ref{sec: Bacterial sublimit}, we can associate
$\alp$ with $\theta$, $\muc$ with $\psi$, and $\phic$ with $\phi$.

\subsection{Solvability conditions from the next-order system}
\label{sec: FC}

Our remaining goal is to determine the governing equations satisfied by the slow-time functions $\alp(\tl)$, $\muc(\tl)$, and $\phic(\tl)$. We note that the procedure to do this is nonstandard for our problem, as it involves a nontrivial system of equations.

Specifically, when using the method of multiple scales for standard problems involving first-order ODEs, procedures to obtain governing equations for the slow-time functions often involve solving the next-order system of the general form
\begin{align}
\label{eq: O eps system}
L \bs{x}_1 = \bs{F}(\bs{x}_0, \partial \bs{x}_{0}/\partial \tl),
\end{align}
where $\bs{x}_1$ is periodic in $\ts$ (say $2 \pi$-periodic, without loss of generality), $L$ is a differential-algebraic linear operator (generally dependent on $\bs{x}_0(\ts,\tl)$), and $\bs{F}$ is a function of both $\bs{x}_0$ and its slow-time derivative. Then one proceeds by visually identifying secular terms in the solution and forcing their coefficients to vanish, generating the appropriate slow-time governing equations \citep{hinch_1991,Bender1999}. However, this procedure can be challenging for more complicated systems where full solutions of the next-order system are prohibitively difficult to obtain, or where secular terms are difficult to identify.

In such systems, it can often be simpler to obtain the requisite governing equations by deriving solvability conditions for the next-order system instead, through the imposition of periodicity on the fast scale. This can be simpler because it circumvents the need to solve an inhomogeneous problem. The appropriate way to derive these solvability conditions is via the method of multiple scales for systems (see, for example, pp.~127--128 of \citet{dalwadi2014flow} or p.~22 of \citet{dalwadi2018effect}), which follows from the Fredholm Alternative Theorem applied to a periodic system. To briefly summarise here: a next-order linear system such as \eqref{eq: O eps system} yields appropriate slow-time evolution equations via its solvability condition
\begin{align}
\label{eq: general solv cond}
\int_0^{2\pi} \! \bsX \bcdot \bs{F} \, \mathrm{d}T = 0,
\end{align}
where $\bsX$ is the fast-time-periodic solution of the homogeneous adjoint problem
\begin{align}
\label{eq: adjoint system}
L^* \bsX = \vec{0},
\end{align}
with fast-time periodicity conditions, where $L^*$ is the differential-algebraic linear adjoint operator. The solvability conditions \eqref{eq: general solv cond} will generate the appropriate slow-time governing equations, reducing our task to explicitly solving the \emph{homogeneous} 3D system \eqref{eq: adjoint system} instead of the \emph{inhomogeneous} 3D system \eqref{eq: O eps system}. Generally, each linearly independent solution of the adjoint problem \eqref{eq: adjoint system} will contribute one solvability condition.

For the specific problem we consider, the next-order system \eqref{eq: O eps system} constitutes a 3D dynamical system for $\bs{x}_1(\ts,\tl) \coloneqq (\theta_1, \psi_1, \phi_1)^{T}$, and is generated from the $\order{1}$ terms in \eqref{eq: full gov eq trans General} after posing the asymptotic expansions \eqref{eq: asy exp bacterial}. Specifically, the linear operator $L$ and the right-hand side $\bs{F}$ are
\begin{align}
\label{eq: L and F def}
L &= 
\begin{pmatrix}
\om \partial_{\ts} & \ac \sin \psi_0 & 0\\
-\ac \dfrac{\sin \psi_0}{\sin^2 \theta_0} & \om \partial_{\ts} + \ac \dfrac{\cos \theta_0 \cos \psi_0}{\sin \theta_0} & 0 \\
\ac\dfrac{\cos \theta_0 \sin \psi_0}{\sin^2 \theta_0} & -\ac\dfrac{\cos \psi_0}{\sin \theta_0} & \om \partial_{\ts}
\end{pmatrix},
\\
\label{eq: F def}
\bs{F} &= 
\begin{pmatrix}
\fb(\theta_0,\phi_0) - \partial \theta_0 / \partial \tl \\
\fc(\theta_0,\phi_0) -  \partial \psi_0 / \partial \tl \\
\fa(\phi_0) - \partial \phi_0 / \partial \tl
\end{pmatrix},
\end{align}
where $\partial_{\ts} \equiv \partial/\partial \ts$. The homogeneous adjoint operator $L^*$ is the transpose of the matrix operator taking the adjoint of each element, and is therefore defined as
\begin{align}
\label{eq: adjoint matrix}
L^* = 
\begin{pmatrix}
 -\om \partial_{\ts} & -\ac\dfrac{\sin \psi_0}{\sin^2 \theta_0} & \ac\dfrac{\cos \theta_0 \sin \psi_0}{\sin^2 \theta_0} \\
 \ac \sin \psi_0 & -\om \partial_{\ts} + \ac \dfrac{\cos \theta_0 \cos \psi_0}{\sin \theta_0} & -\ac\dfrac{\cos \psi_0}{\sin \theta_0} \\
0 & 0 & -\om \partial_{\ts} \\
\end{pmatrix}.
\end{align}
We emphasise that $L \neq L^*$ i.e. this problem is \emph{not} self-adjoint.

To generate the slow-time equations we are seeking for the three slow-time functions $\alp(\tl)$, $\muc(\tl)$, and $\phic(\tl)$, we must derive the appropriate solvability conditions \eqref{eq: general solv cond} by solving the homogeneous adjoint problem \eqref{eq: adjoint system} with $L^*$ defined in \eqref{eq: adjoint matrix}. Given that the coefficients in \eqref{eq: adjoint matrix} depend on the fast time $\ts$, this is a nontrivial task. Nevertheless, we are able to solve \eqref{eq: adjoint system} in Appendix~\ref{sec: next order adjoint}, where we deduce that the general solution is
\begin{align}
\bsX 
&= \Cb
\begin{pmatrix}
\ac \cos \theta_0 \sin \psi_0 - \sin \theta_0  \\ \ac \sin \theta_0 \cos \psi_0 \\ 0
\end{pmatrix}
\notag \\
\label{eq: adjoint solution}
&\quad
+ \Cc
\begin{pmatrix}
\ac \cos \psi_0 \\ \sin \theta_0 \left( \sin \theta_0 - \ac \cos \theta_0 \sin \psi_0\right) \\ 0
\end{pmatrix}
+ \Ca
\begin{pmatrix}
0 \\ \cos \theta_0 \\ 1
\end{pmatrix}
,
\end{align}
for arbitrary constants $\Cb$, $\Cc$, and $\Ca$.

Therefore, we obtain our required solvability conditions by substituting the adjoint solutions \eqref{eq: adjoint solution} into the general solvability condition \eqref{eq: general solv cond}, with $\bs{F}$ defined in \eqref{eq: F def}, and setting the resulting coefficients of $\Ci$ ($i = 1, 2, 3$) to zero. This procedure yields the following three solvability conditions
\begin{subequations}
\label{eq: solv conditions orig}
\begin{align}
&\av{\theta_{0 \tl} \left(\ac \cos \theta_0 \sin \psi_0 - \sin \theta_0 \right) + \psi_{0 \tl} \ac \sin \theta_0 \cos \psi_0} \notag \\
\label{eq: solv conditions orig fb fc 1}
&\qquad= \av{\fb \left(\ac \cos \theta_0 \sin \psi_0 - \sin \theta_0 \right) +  \fc \ac \sin \theta_0 \cos \psi_0}, \\
&\av{\theta_{0 \tl} \ac \cos \psi_0 + \psi_{0 \tl} \sin \theta_0 \left(\sin \theta_0 - \ac \cos \theta_0 \sin \psi_0 \right)} \notag \\
\label{eq: solv conditions orig fb fc 2}
&\qquad= \av{\fb \ac \cos \psi_0 + \fc \sin \theta_0 \left( \sin \theta_0 - \ac \cos \theta_0 \sin \psi_0 \right)}, \\
\label{eq: solv conditions orig fa fc}
&\av{\psi_{0\tl} \cos \theta_0 + \phi_{0\tl}} = \av{\fc \cos \theta_0 + \fa},
\end{align}
\end{subequations}
where we use the subscript $\tl$ to denote partial differentiation with respect to $\tl$, and we use the notation $\av{\bcdot}$ to denote the average of its argument over one fast-time oscillation, as defined in \eqref{eq: av operator}.

Our remaining task is to evaluate the averages in \eqref{eq: solv conditions orig} in terms of the slow-term functions $\alp$, $\muc$, and $\phic$, and this will lead us to the emergent slow-time equations for the system. The left-hand sides of \eqref{eq: solv conditions orig} are currently written in terms of the original variables $\theta_0(\ts,\tl)$, $\psi_0(\ts,\tl)$, and $\phi_0(\ts,\tl)$. To proceed, we need to write them in terms of the slow-time functions $\alp(\tl)$, $\muc(\tl)$, and $\phic(\tl)$ and then take appropriate averages. We carry out this task in Appendix \ref{sec: LHS of solv cond}, which allows us to rewrite \eqref{eq: solv conditions orig} as
\begin{subequations}
\label{eq: solv conditions transformed}
\begin{align}
- \om \sin \alp \dbyd{\alp}{\tl} &= \av{\fb \left(\ac \cos \theta_0 \sin \psi_0 - \sin \theta_0 \right) + \fc \ac \sin \theta_0 \cos \psi_0}, \\
\om \sin^2 \alp \dbyd{\muc}{\tl} &= \av{\fb \ac \cos \psi_0 + \fc \sin \theta_0 \left( \sin \theta_0 - \ac \cos \theta_0 \sin \psi_0 \right)}, \\
\cos \alp \dbyd{\muc}{\tl} + \dbyd{\phic}{\tl} &= \av{\fc \cos \theta_0 + \fa}.
\end{align}
\end{subequations}
The equations \eqref{eq: solv conditions transformed} represent the appropriate solvability conditions for \emph{any} $f_i$ acting on a rapidly spinning object in Stokes flow. In the remainder of Part 1, we evaluate these solvability conditions for the $f_i$ defined in \eqref{eq: f functions}, which represent the hydrodynamic interaction of shear flow with a spheroidal object. In Part 2, we consider the interaction of shear flow with general chiral objects with helicoidal symmetry. This requires evaluating a different set of functions on the right-hand side of \eqref{eq: solv conditions transformed}, resulting in fundamentally different emergent equations to Part 1.

\subsection{Evaluating the right-hand side of the solvability conditions}

Our final task is to evaluate the right-hand side of the three solvability conditions in \eqref{eq: solv conditions transformed}, using the $f_i$ defined in \eqref{eq: f functions}. While \eqref{eq: mu def} provide helpful expressions for the required trigonometric functions of $\theta_0$ and $\psi_0$, currently our only expression for $\phi_0$ is \eqref{eq: tan phi0}, given in terms of $\tan \phi_0$. Given that the $f_i$ defined in \eqref{eq: f functions} involve $\sin 2 \phi_0$ and $\cos 2 \phi_0$, it is helpful to derive expressions for these quantities. To do this, we first calculate the following expressions involving $\sin \phi_0$ and $\cos \phi_0$ from $\tan \phi_0$ defined in \eqref{eq: tan phi0}:
\begin{subequations}
\label{eq: a 0 LO identities General a}
\begin{align}
\label{eq: a 0 LO identities General a sin thet sin phi}
\om \sin \theta_0 \sin \phi_0 &= \ac \cos \phic \sin \sig + \sin \phic \left(\ac \cos \alp \cos \sig + \sin \alp \right), \\
\label{eq: a 0 LO identities General a sin thet cos phi}
\om \sin \theta_0 \cos \phi_0 &= \cos \phic \left(\ac \cos \alp \cos \sig + \sin \alp \right) - \ac \sin \phic \sin \sig,
\end{align}
\end{subequations}
using the shorthand notation
\begin{align}
\label{eq: sigma def}
\sig = \ts + \muc(\tl),
\end{align}
for algebraic convenience. The factors of $\om \sin \theta_0$ in 
\eqref{eq: a 0 LO identities General a} arise from imposing the constraint $\sin^2 \phi_0 + \cos^2 \phi_0 = 1$ and using \eqref{eq: om sin thet sq def} to evaluate $\sin^2 \theta_0$. While it is not essential, since the sign differences would cancel with one another in our subsequent analysis, for definiteness we choose the positive square root, essentially exploiting the fact that $\sin \theta_0 > 0$ (since $\theta_0 \in [0, \pi]$), and using our choice of $\phic(0)$ up to an additive multiple of $\pi$ (see Appendix \ref{sec: Slow time functions IC}). Then, from the expressions \eqref{eq: a 0 LO identities General a}, we may use appropriate double-angle formulae to deduce that
\begin{subequations}
\begin{align}
\label{eq: cos 2 phi}
\om^2 \sin^2 \theta_0 \cos 2 \phi_0 &= \Ceven(\sig,\tl) \cos 2\phic - \Sodd(\sig,\tl) \sin 2 \phic, \\
\label{eq: sin 2 phi}
\om^2 \sin^2 \theta_0 \sin 2 \phi_0 &= \Sodd(\sig,\tl) \cos 2\phic + \Ceven(\sig,\tl) \sin 2 \phic,
\end{align}
where we have introduced the functions
\begin{align}
\label{eq: Ceven and Sodd}
\Ceven(\sig,\tl) &\coloneqq \left(\ac \cos \alp \cos \sig + \sin \alp\right)^2 - \ac^2 \sin^2\sig, \\
\label{eq: Sodd}
\Sodd(\sig,\tl) &\coloneqq 2 \ac \sin\sig \left(\ac \cos \alp \cos \sig + \sin \alp \right).
\end{align}
\end{subequations}

To evaluate the fast-time averages in the right-hand side of \eqref{eq: solv conditions transformed}, it is helpful to exploit the parity of various quantities. In particular, we note that $\Ceven$ and $\Sodd$ are even and odd, respectively, around $\sig = \pi$, which follows immediately from their definitions \eqref{eq: Ceven and Sodd}--\eqref{eq: Sodd}. Since $\sin^2 \theta_0$ is also even, as follows from \eqref{eq: om sin thet sq def}, this means that $\cos 2 \phi_0$ and $\sin 2 \phi_0$ defined in \eqref{eq: cos 2 phi}--\eqref{eq: sin 2 phi} are naturally decomposed into their odd and even parts. Additionally, since the expressions in \eqref{eq: mu def} tell us that $\cos \theta_0$ and $\sin \theta_0 \sin \psi_0$ are even, and that $\sin \theta_0 \cos \psi_0$ is odd (each around $\sig = \pi$), several terms on the right-hand side of \eqref{eq: solv conditions transformed} will vanish immediately under fast-time averaging. Exploiting this parity, we find that
\begin{subequations}
\label{eq: non chiral in theta psi form}
\begin{align}
\label{eq: alp eq no chiral app}
&\av{\fb \left(\ac \cos \theta_0 \sin \psi_0 - \sin \theta_0 \right) + \fc \ac \sin \theta_0 \cos \psi_0} \notag \\
 & \qquad = \dfrac{\Breth}{2 \om^2}\sin 2 \phic \av{\cos \theta_0 \left[\Ceven \left(1 - \dfrac{\ac \cos \theta_0 \sin \psi_0}{\sin \theta_0}\right)  - \dfrac{\Sodd \ac \cos \psi_0}{\sin \theta_0} \right]}, \\
\label{eq: mu eq no chiral app}
&\av{\fb \ac \cos \psi_0 + \fc \sin \theta_0 \left( \sin \theta_0 - \ac \cos \theta_0 \sin \psi_0 \right)} \notag \\
& \qquad = \dfrac{\Breth}{2 \om^2}\cos 2 \phic \av{\cos \theta_0 \left[\Ceven \left(1 - \dfrac{\ac \cos \theta_0 \sin \psi_0}{\sin \theta_0}\right)  - \dfrac{\Sodd \ac \cos \psi_0}{\sin \theta_0} \right]}, \\
\label{eq: phi eq no chiral app}
&\av{\fc \cos \theta_0 + \fa} = \dfrac{1}{2}\left(1 - \dfrac{\Breth}{\om^2} \av{\Ceven} \cos 2 \phic_0 \right).
\end{align}
\end{subequations}

To evaluate the required average in \eqref{eq: alp eq no chiral app} and \eqref{eq: mu eq no chiral app} (which is the same), we use the relationships \eqref{eq: mu def}, \eqref{eq: Ceven and Sodd}--\eqref{eq: Sodd}, and simplify the resulting expressions to deduce
\begin{subequations}
\label{eq: requisite averages}
\begin{align}
&\av{\cos \theta_0 \left[\Ceven \left(1 - \dfrac{\ac \cos \theta_0 \sin \psi_0}{\sin \theta_0}\right)  - \dfrac{\Sodd \ac \cos \psi_0}{\sin \theta_0} \right]} \notag \\
& \qquad = \dfrac{\om \sin \alp}{2}  \av{2 \ac \cos \sig \cos 2 \alp + (2 - \ac^2(1 + \cos 2\sig ) ) \sin \alp \cos \alp} \notag \\
\label{eq: cos theta sin theta average}
& \qquad = \dfrac{(2 - \ac^2 ) \om \sin^2 \alp \cos \alp}{2}.
\end{align}
Similarly, the required average in \eqref{eq: phi eq no chiral app} can be evaluated using \eqref{eq: om sin thet sq def} and \eqref{eq: Ceven and Sodd}--\eqref{eq: Sodd}, to obtain
\begin{align}
\label{eq: CC^2 - SS^2}
\av{\Ceven} = \av{\left(\ac \cos \alp \cos \sig + \sin \alp\right)^2 - \ac^2 \sin^2\sig} = \dfrac{2 - \ac^2}{2} \sin^2 \alp.
\end{align}
\end{subequations}
Then using the results \eqref{eq: requisite averages} in \eqref{eq: non chiral in theta psi form}, we can write
\begin{subequations}
\label{eq: g_i RHS}
\begin{align}
\av{\fb \left(\ac \cos \theta_0 \sin \psi_0 - \sin \theta_0 \right) + \fc \ac \sin \theta_0 \cos \psi_0} &= \dfrac{2 - \ac^2}{4 \om} \Breth \sin^2 \alp \cos \alp \sin 2 \phic, \\
\av{\fb \ac \cos \psi_0 + \fc \sin \theta_0 \left( \sin \theta_0 - \ac \cos \theta_0 \sin \psi_0 \right)} &=  \dfrac{2 - \ac^2}{4 \om} \Breth \sin^2 \alp \cos \alp \cos 2 \phic, \\
\av{\fc \cos \theta_0 + \fa} &= \dfrac{1}{2}\left(1 - \dfrac{2 - \ac^2}{2 \om^2}\Breth \sin^2 \alp \cos 2 \phic \right).
\end{align}
\end{subequations}
\Cref{eq: g_i RHS} concludes our evaluation of the right-hand sides of the solvability conditions. The next step is to piece this all together to obtain the emergent angular dynamics.

\subsection{The emergent angular dynamics}

We now have the requisite information to obtain the governing equations for $\alp$, $\muc$, and $\phic$, which are the main results we have been seeking through this analysis. To this end, we substitute the right-hand side results \eqref{eq: g_i RHS} into the solvability conditions \eqref{eq: solv conditions transformed}, then rearrange to obtain the following nonlinear system
\begin{subequations}
\label{eq: original reduced}
\begin{align}
\dbyd{\alp}{\tl} &= -\dfrac{\Breth}{4} \dfrac{2 - \ac^2}{1 + \ac^2}\sin \alp \cos \alp \sin 2 \phic, \\
\dbyd{\muc}{\tl} &= \dfrac{\Breth}{4} \dfrac{2 - \ac^2}{1 + \ac^2} \cos \alp \cos 2 \phic, \\
\dbyd{\phic}{\tl} &= \dfrac{1}{2}\left(1 - \dfrac{2 - \ac^2}{2 \left(1 + \ac^2 \right)}\Breth \cos 2 \phic \right),
\end{align}
\end{subequations}
where $\alp$, $\muc$, and $\phic$ feed into the full leading-order asymptotic solution through \eqref{eq: mu def} and \eqref{eq: tan phi0}.

Remarkably, \eqref{eq: original reduced} can be re-written in terms of the functions $f_i$, which are defined in \eqref{eq: f functions}, to yield
\begin{subequations}
\label{eq: simplified reduced}
\begin{align}
\dbyd{\alp}{\tl} = \fb(\alp,\phic; \Beff), \quad
\dbyd{\muc}{\tl} = \fc(\alp,\phic;  \Beff), \quad
\dbyd{\phic}{\tl} = \fa(\phic; \Beff),
\end{align}
where we define $\Beff$ to be the effective Bretherton parameter, given explicitly by
\begin{align}
\label{eq: effective coefficients}
\Beff \coloneqq \dfrac{2 - \ac^2}{2 \left(1 + \ac^2 \right)}\Breth.
\end{align}
\end{subequations}

\begin{figure}
    \centering
    \medskip
    \begin{overpic}[width=\textwidth]{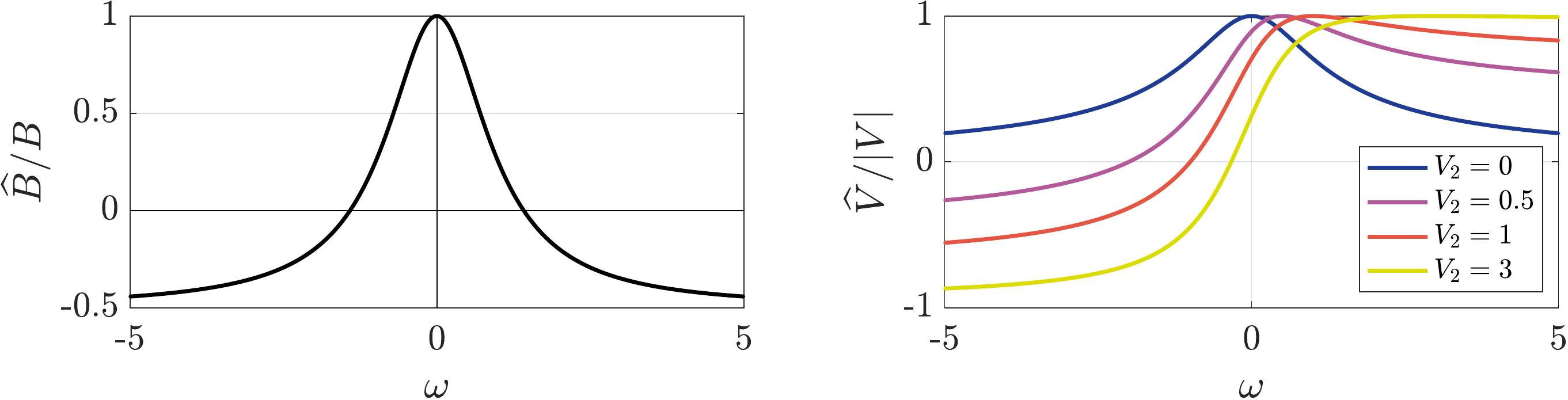}
    \put(1,24){(a)}
    \put(53,24){(b)}
    \end{overpic}
    \caption{The effective parameters as a function of $\ac$: (a) $\Beff$ relative to $\Breth$ from \eqref{eq: effective coefficients}, and (b) $\veleff$ relative to $|V|  = \sqrt{\velscala^2 + \velscalb^2}$ from \eqref{fin0}. Of note, reversing the sign of $\velscalb$ would reflect each line in the $\ac = 0$ axis.}
    \label{fig:B_hat_V_hat}
\end{figure}

Writing the system \eqref{eq: original reduced} in the form \eqref{eq: simplified reduced} leads to several key deductions. Firstly, by comparison with the full system \eqref{eq: full gov eq}--\eqref{eq: f functions}, we see that we can formally identify each slow-time function with an underlying variable: $\alp$ with $\theta$, $\muc$ with $\psi$, and $\phic$ with $\phi$. Moreover, the compact relationship described by \eqref{eq: simplified reduced} retroactively motivates our notational choice to define the effective fast-time conserved quantity as $\om \cos \alp(\tl)$ in \eqref{eq: alpha def} instead of $H(\tl)$ as in \eqref{eq: H with H}.

Secondly, the effective Bretherton parameter is given in \eqref{eq: effective coefficients}, and plotted with respect to $\ac$ on \Cref{fig:B_hat_V_hat}a, from which we note that the sign of the effective Bretherton parameter is different to that of the actual Bretherton parameter if $|\ac| > \sqrt{2}$. Moreover, the effective Bretherton parameter vanishes (i.e. the effective particle behaves as a sphere) if $|\ac| = \sqrt{2}$. More generally, since \eqref{eq: simplified reduced} is exactly equivalent to Jeffery's equations, we have found that the rapid spinning of spheroidal swimmers generates Jeffery's orbits with an effective Bretherton parameter that we have calculated analytically. Therefore, the effect of rapid spinning is to modify the effective shape of the spheroidal swimmer, as quantified through the relationship \eqref{eq: effective coefficients}.

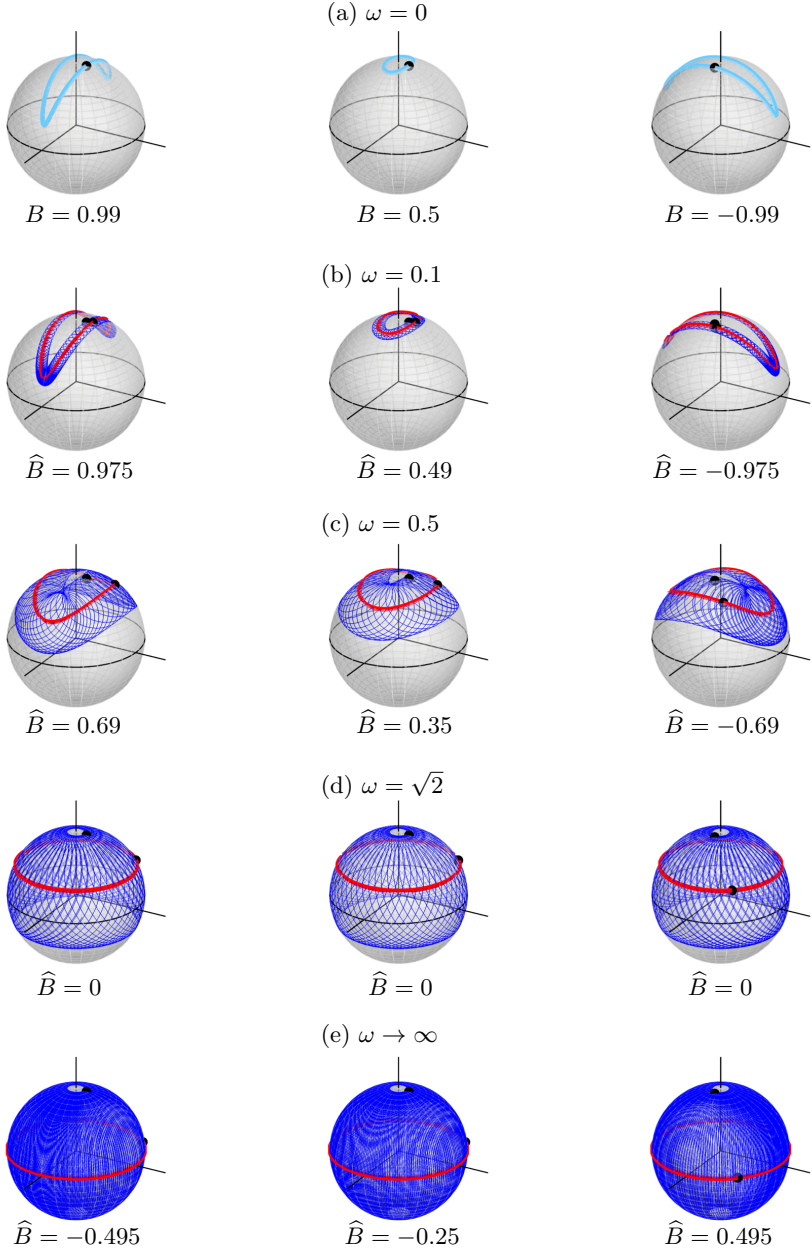
\begin{figure}
    \centering
    \vspace{2em}
    
    \begin{overpic}[width=0.82\textwidth]{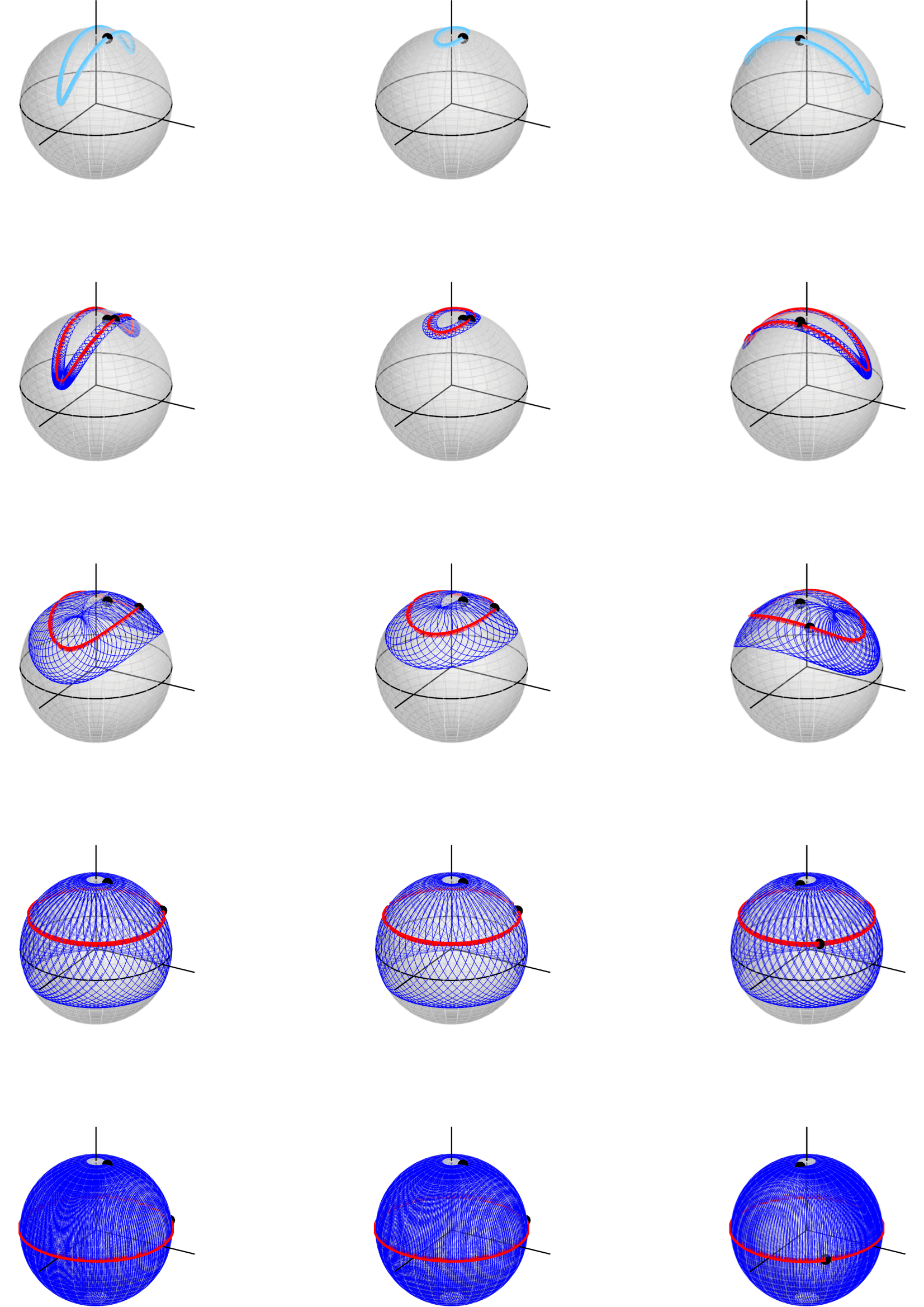}
    \put(28.5,101){(a) $\ac = 0$}
    \put(3,84){$\Breth = 0.99$}
    \put(31,84){$\Breth = 0.5$}
    \put(57,84){$\Breth = -0.99$}
    \put(28,79){(b) $\ac = 0.1$}
    \put(3,62.5){$\Beff = 0.975$}
    \put(31,62.5){$\Beff = 0.49$}
    \put(56,62.5){$\Beff = -0.975$}
    \put(28,58){(c) $\ac = 0.5$}
    \put(3,41){$\Beff = 0.69$}
    \put(31,41){$\Beff = 0.35$}
    \put(57,41){$\Beff = -0.69$}
    \put(28,36){(d) $\ac = \sqrt{2}$}
    \put(4,19){$\Beff = 0$}
    \put(32,19){$\Beff = 0$}
    \put(59,19){$\Beff = 0$}
    \put(28,15){(e) $\ac \rightarrow \infty$}
    \put(2,-2){$\Beff = -0.495$}
    \put(30,-2){$\Beff = -0.25$}
    \put(57,-2){$\Beff = 0.495$}
    \end{overpic}
    
    \vspace{1em}
    \centering
    \caption{Rotational dynamics of a spheroidal swimmer with general axis of rotation. Each plot displays the evolution of the particle orientation $(\theta,\phi)$, with vertical axis corresponding to $\theta=0$. In each column, we fix $\Breth$ and vary $\ac$, so each column corresponds to the same object shape. In rows (b) to (e), the full evolution of the particle orientation \eqref{eq: full gov eq}--\eqref{eq: f functions} is shown as a thin blue line, while the emergent dynamics \eqref{eq: simplified reduced} are shown as a thick red line. (a) Jeffery's orbits of passive particles, for reference. (b) The bacterial limit, with $\abs{\ac}$ small. (c)--(e). Larger values of $\ac$ yield Jeffery's orbits with an effective Bretherton parameter for the emergent dynamics. (d) The critical value $\ac =\sqrt{2}$ leads to $\Beff=0$ for each particle. (e) $\Breth$ and $\Beff$ having opposite signs for $\abs{\ac} > \sqrt{2}$. We use $\omb = 15$ and initial conditions $\theta(0) = \pi/18$, $\psi(0) = -\pi/4$ throughout. We use initial condition $\phi(0) = \pi/2$ for the first two columns, and $\phi(0) = 0$ for the final column. The initial condition for each trajectory is shown as a black dot.}
    \label{fig:rotational_GD}
\end{figure}

Finally, it is of interest to briefly consider the limiting cases of $\ac \to 0$ and $|\ac|  \to \infty$. We have already considered the limit of $\ac \to 0$ (where the rotation axis coincides with the spheroidal symmetry axis) explicitly in \S \ref{sec: Bacterial sublimit}. It is reassuring to note that in this limit, our generalised effective Bretherton parameter $\Beff \to \Breth$ from \eqref{eq: effective coefficients}, and hence the system \eqref{eq: simplified reduced} reduces to
\begin{align}
\label{eq: small a reduced}
\dbyd{\alp}{\tl} = \fb(\alp,\phic; \Breth), \quad
\dbyd{\muc}{\tl} = \fc(\alp,\phic; \Breth), \quad
\dbyd{\phic}{\tl} = \fa\left(\phic; \Breth \right),
\end{align}
in agreement. That is, this type of spinning leaves the system essentially unchanged, recapturing the explicit results of \S \ref{sec: Bacterial sublimit}.

In the limit of $|\ac|  \to \infty$ (where the rotation axis is perpendicular to the spheroidal symmetry axis), our generalised effective Bretherton parameter $\Beff \to -\Breth/2$ from \eqref{eq: effective coefficients}, and the system \eqref{eq: simplified reduced} reduces to
\begin{align}
\label{eq: large a reduced}
\dbyd{\alp}{\tl} = \fb\left(\alp,\phic; -\frac{\Breth}{2}\right), \quad
\dbyd{\muc}{\tl} = \fc\left(\alp,\phic; -\frac{\Breth}{2}\right), \quad
\dbyd{\phic}{\tl} = \fa\left(\phic;  -\frac{\Breth}{2} \right).
\end{align}
That is, the sign of the effective Bretherton parameter changes (so an oblate spheroid behaves as a prolate spheroid, and vice versa) and its magnitude halves (making it behave more like a sphere, in some sense).

We illustrate our asymptotic results in \Cref{fig:rotational_GD}, showing the rotational dynamics of the swimmer orientation $(\theta,\phi)$ via the same representation on the unit sphere as in \Cref{fig:rotational_BL}. With \Cref{fig:rotational_GD}a showing the Jeffery's orbits of passive particles for reference, it is clear that rapid spinning with non-zero $\ac$ can significantly impact on the emergent dynamics, with $\ac$ and the Bretherton parameter $\Breth$ varying between panels. Within each column, the initial configuration is fixed and shown as a black dot for reference. The average trajectory predicted by the asymptotic analysis is shown as a red curve in each plot, and the full numerical solution is shown as a blue curve. In each case, the full asymptotic solution obtained from combining the leading-order fast-time solutions \eqref{eq: mu def} and \eqref{eq: tan phi0} with the slow-time equations \eqref{eq: simplified reduced} successfully captures the leading-order behaviour of the full oscillatory dynamics, even when it is markedly complex, though we omit this comparison in \Cref{fig:rotational_GD} to allow the details of the trajectories to be seen cleanly.

As predicted by equation \eqref{eq: simplified reduced}, the emergent dynamics of the spinning objects are simply Jeffery's orbits in transformed variables with modified Bretherton parameters. That is, a rapidly spinning object behaves as though it is a differently shaped spin-free object. In particular, with $\abs{\Beff} < \abs{\Breth}$, the emergent Jeffery's orbits are somewhat simpler, in that they more closely resemble the straightforward behaviour exhibited by a sphere. This is particularly evident as $\abs{\ac}\rightarrow\sqrt{2}$, with the dynamics approaching those of a sphere for \emph{all} particles irrespective of their shape, as shown in \Cref{fig:rotational_GD}d.

Our final task is to use our results for the emergent angular dynamics to calculate the emergent translational dynamics.

\subsection{The emergent translational dynamics}\label{etd}

In this final subsection for the general problem, we calculate the emergent translational dynamics, for which the full governing equations are given in \eqref{eq: full gov eq translational}. Our aim is to derive emergent (effective) governing equations for the translation, again using the method of multiple scales. 
Using the time derivative transformation \eqref{eq: time deriv transform}, the governing equations \eqref{eq: full gov eq translational} become
\begin{multline}
\label{eq: transform gov eq translational}
    \omb \om \pbyp{\Xvecpos}{\ts} + \pbyp{\Xvecpos}{\tl} = \velscala\ehat{1}(\theta, \phi) + \velscalb\ehat{2}(\theta,\psi, \phi) + \velscalc\ehat{3}(\theta,\psi, \phi) + \Ypos \e{3}.
\end{multline}

We proceed as before, taking an asymptotic expansion of the dependent variables in inverse powers of $\omb$, as in \eqref{eq: asy exp bacterial}. At leading order (i.e. $\mathit{O}(\omb)$), we obtain
\begin{align}
\label{eq: transform gov eq translational LO general}
\pbyp{\Xvecpos_0}{\ts} = 0.
\end{align}
Therefore, in the same way as for the bacterial sublimit in \S \ref{sec: Bacterial sublimit}, $\Xvecpos_0 = \Xvecpos_0(\tl)$ and the vector position $\Xvecpos$ is independent of the fast time $\ts$ at leading order.

At next order (i.e. $\mathit{O}(1)$), substituting the leading-order fast-time angular dynamics solutions \eqref{eq: mu def} and \eqref{eq: tan phi0} in \eqref{eq: transform gov eq translational}, we obtain the system
\begin{multline}
\label{eq: transform gov eq translational FC}
\om \pbyp{\Xvecpos_1}{\ts} + \dbyd{\Xvecpos_0}{\tl} = \velscala\ehat{1}(\theta_0, \phi_0) + \velscalb\ehat{2}(\theta_0,\psi_0, \phi_0) + \velscalc\ehat{3}(\theta_0,\psi_0, \phi_0) + \Ypos_0 \e{3},
\end{multline}
with a fast-time periodicity constraint on $\Xvecpos_1$. The solvability conditions are straightforward to obtain by integrating \eqref{eq: transform gov eq translational FC} with respect to the fast time $\ts$ from $0$ to $2 \pi$. As before, the fast-time derivative vanishes due to periodicity. 
In Appendix \ref{sec: fast time average of swimmer basis}, we calculate the fast-time averages of the unit vectors as
\begin{equation}
\label{eq: average basis vectors}
    \av{\om \ehat{1}(\theta_0, \phi_0)} = \etilde{1}(\alp,\phic), \quad
    \av{\om \ehat{2}(\theta_0,\psi_0, \phi_0)} = \ac \etilde{1}(\alp,\phic), \quad
    \av{\ehat{3}(\theta_0,\psi_0, \phi_0)} = \vec{0},
\end{equation}
where $\etilde{1}(\alp,\phic)$ can be considered equivalent to the (hatted) basis vector $\ehat{1}$ in \eqref{elm2}, but with argument $(\theta,\phi)$ replaced by $(\alp,\phic)$.

Hence, substituting \eqref{eq: average basis vectors} into the fast-time averaged version of \eqref{eq: transform gov eq translational FC}, the emergent governing equation for translation is
\begin{align}
\label{fin0}
    \dbyd{\Xvecpos_0}{\tl} = \veleff \etilde{1}(\alp,\phic) + \Ypos_0 \e{3}, \qquad \text{with }
    \veleff = \dfrac{\velscala + \ac \velscalb}{\om} = \dfrac{\velscala + \ac \velscalb}{\sqrt{1 + \ac^2}}.
\end{align} 
Of note, and generalising the case explored in \S \ref{sec: Bacterial sublimit}, we now retain a contribution from an off-axis direction of swimming, namely $\velscalb$. That is, this off-axis swimming component does not average to zero in the general case. The evolution of the effective swimming velocity $\veleff$ (normalised by $|V| = \sqrt{\velscala^2+\velscalb^2}$) with respect to $\ac$ is shown in \Cref{fig:B_hat_V_hat}(b) for different values of off-axis velocity. Perhaps surprisingly, the coefficient of $\velscalb$ in \eqref{fin0} is odd in $\ac$, suggesting that the behaviour of a swimmer with $\ac<0$ can be markedly different to that of one with $\ac>0$, recalling that the sign of $\ac$ is determined solely by the relative directions of spinning via \eqref{eq: def omega as ratio of Omegas}. Thus, supposing that $\velscala=\abs{\ac}\velscalb$, a swimmer with both $\oma>0$ and $\omb>0$ propels itself at an effective speed of $2\abs{\velscala}/\om$, while swapping either one of the directions of spinning gives an effective swimming speed of zero. The role of $\velscalb$ is further highlighted in the limit $\abs{\ac}\to\infty$, in which the contribution of $\velscala$ vanishes. Hence, in this limit, which might be considered the `opposite' to the bacterial case, the translational dynamics reduce only to on-axis propulsion, modulated by $\velscalb$, and the shape-independent advection by the shear.

\begin{figure}
    \centering
    \begin{overpic}[width=\textwidth]{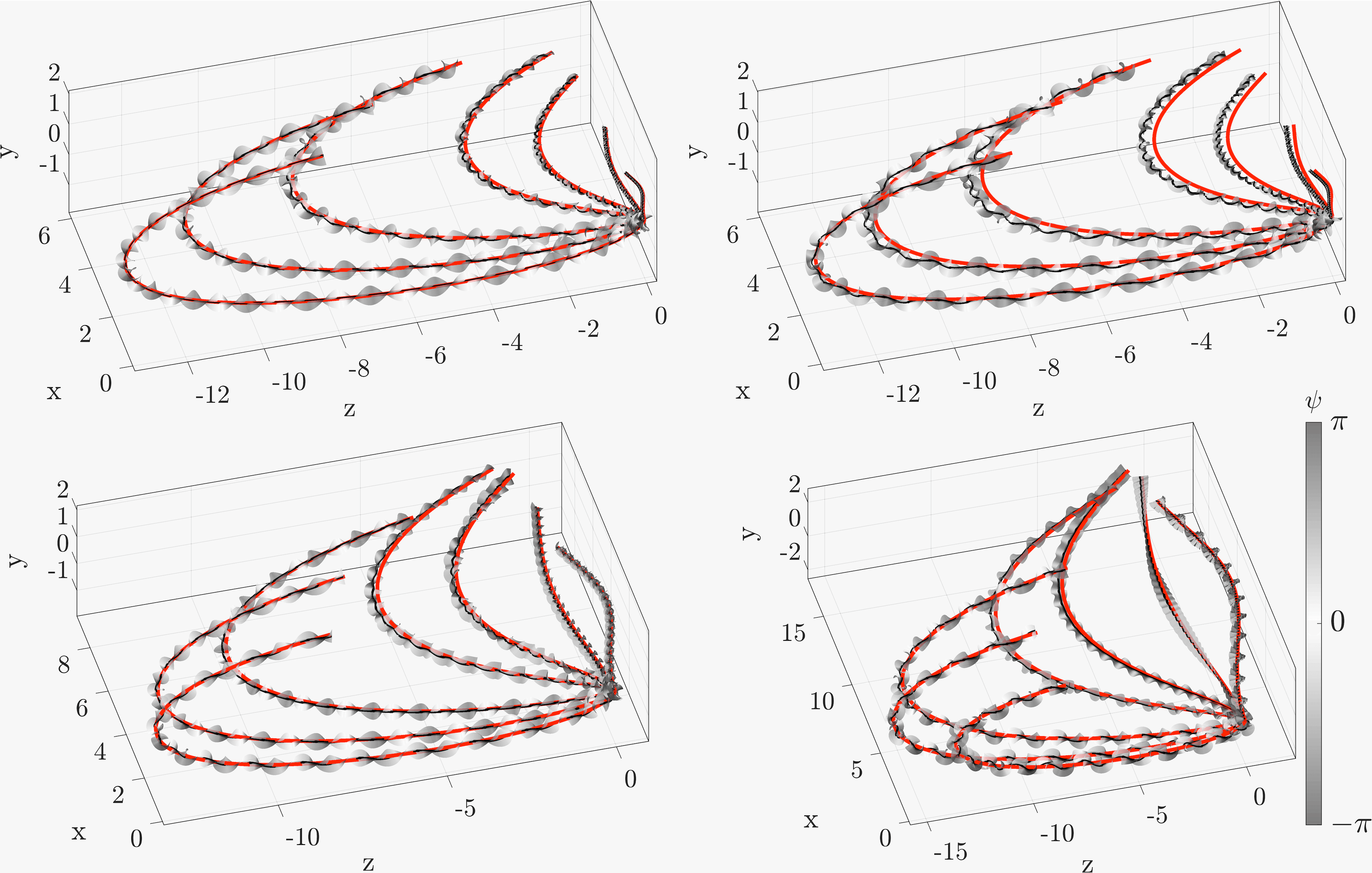}
    \put(0,61){(a)}
    \put(5.5,47.2){$\scriptscriptstyle \ac=0 $}
    \put(9,51.5){$\scriptscriptstyle \ac=0.25$}
    \put(25,59){$\scriptscriptstyle \ac=0.5$}
    \put(37.5,60.5){$\scriptscriptstyle \ac=1$}
    \put(42.3,58.3){$\scriptscriptstyle \ac=\sqrt{2}$}
    \put(42.5,55){$\scriptscriptstyle \ac=2.5$}
    \put(46,51){$\scriptscriptstyle \ac=5$}
    \put(50,61){(b)}
    \put(56,47.2){$\scriptscriptstyle \ac=0 $}
    \put(58.5,51.5){$\scriptscriptstyle \ac=0.25$}
    \put(74.5,59){$\scriptscriptstyle \ac=0.5$}
    \put(86.5,60.5){$\scriptscriptstyle \ac=1$}
    \put(91,59){$\scriptscriptstyle \ac=\sqrt{2}$}
    \put(91.5,55){$\scriptscriptstyle \ac=2.5$}
    \put(95.5,51.5){$\scriptscriptstyle \ac=5$}
    \put(0,30){(c)}
    \put(14,6){$\scriptscriptstyle \ac=0 $}
    \put(7,17){$\scriptscriptstyle \ac=0.25$}
    \put(17,24){$\scriptscriptstyle \ac=0.5$}
    \put(30,29){$\scriptscriptstyle \ac=1$}
    \put(37,30){$\scriptscriptstyle \ac=\sqrt{2}$}
    \put(40,26.5){$\scriptscriptstyle \ac=2.5$}
    \put(43,23){$\scriptscriptstyle \ac=5$}
    \put(50,30){(d)}
    \put(70,6){$\scriptscriptstyle \ac=0$}
    \put(58.7,9.5){$\scriptscriptstyle \ac=0.25 $}
    \put(62,18){$\scriptscriptstyle \ac=0.5$}
    \put(71,25){$\scriptscriptstyle \ac=1$}
    \put(80,30){$\scriptscriptstyle \ac=\sqrt{2}$}
    \put(84.2,28){$\scriptscriptstyle \ac=2.5$}
    \put(89,24){$\scriptscriptstyle \ac=5$}
    \end{overpic}
    \caption{Translational dynamics of a spheroidal swimmer with general axis of rotation. A dynamic version of this figure is given in Supplementary Movie 2. The solutions of the full dynamical system \eqref{eq: full gov eq}--\eqref{eq: full gov eq translational} are plotted as ribbons with a black centreline, while the emergent dynamics \eqref{eq: simplified reduced}, \eqref{fin0} are shown as thick red lines that accurately capture the behaviour of the full system. The orthogonal velocities vary for each panel: (a) $(\velscalb,\velscalc) = (0,0)$, for which we see that increasing $\ac$ reduces the effective swimming speed; (b) $(\velscalb,\velscalc) = (0,0.5)$, for which we see that the average dynamics are unaltered from (a), despite significant modifications in the fast-time variation; (c) $(\velscalb,\velscalc) = (0.5,0)$; (d) $(\velscalb,\velscalc) = (1.5,0)$. In (c) and (d), we see that the dynamics are now significantly modified from (a) and (b) (an effect that increases with $\ac$), but are still captured by the asymptotic solution at leading order. In each panel, we use $\Breth = 0.5$, $\omb = 10$, $\velscala = 1$, initial position $\Xvecpos = \vec{0}$, and initial orientation $(\theta,\psi,\phi) = (2\pi/5,\pi/2,-7\pi/15)$, simulating over the same time interval.}
    \label{fig:translational_GD}
\end{figure}

\begin{figure}
    \centering
    \begin{overpic}[width=\textwidth]{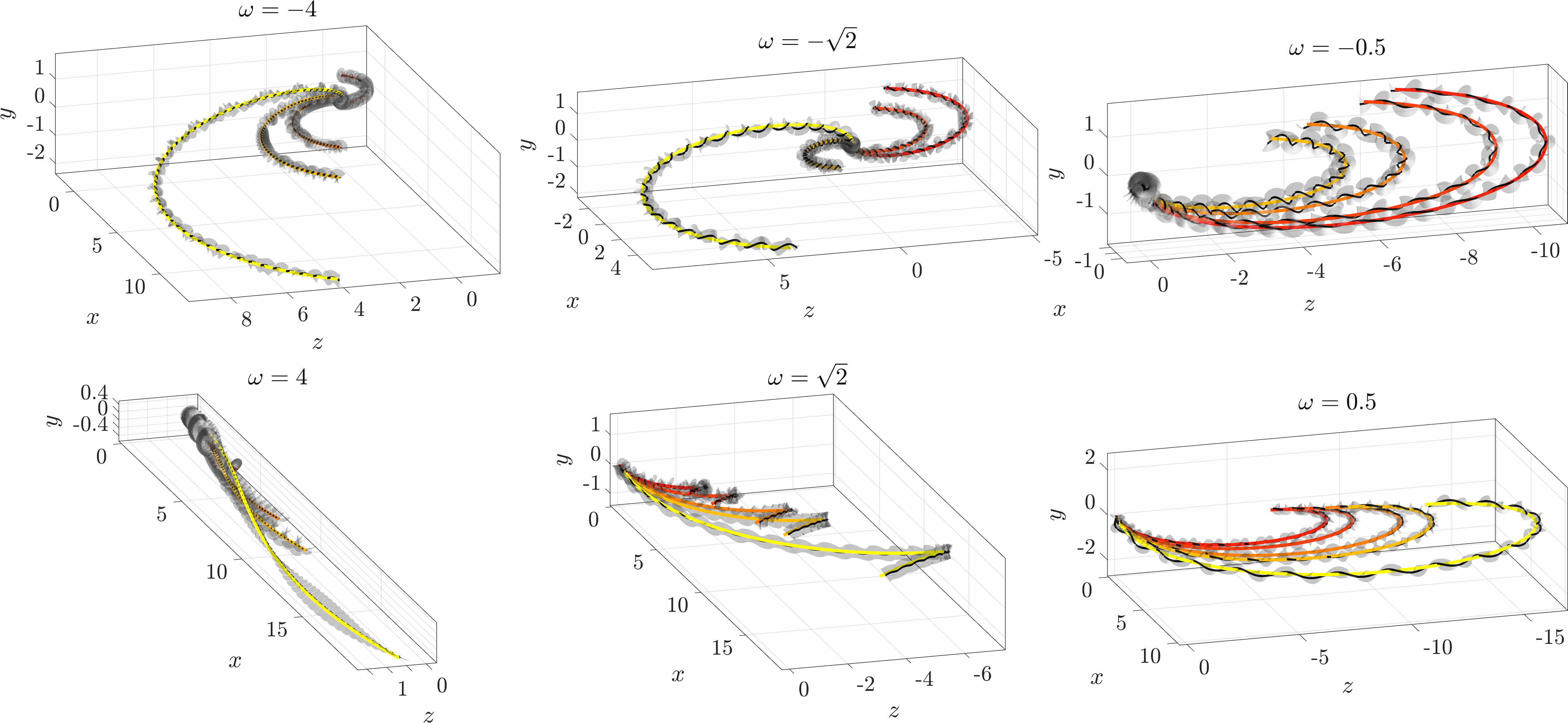}
    \end{overpic}
    \caption{Translational dynamics of a spheroidal swimmer with general axis of rotation, comparing to negative values of $\ac$. In each case, the solutions of the full dynamical system \eqref{eq: full gov eq}--\eqref{eq: full gov eq translational} are plotted as ribbons with a black centreline, while the emergent dynamics \eqref{eq: simplified reduced}, \eqref{fin0} are shown as thick coloured lines that accurately capture the behaviour of the full system. The colour of the thick lines indicates the value of the orthogonal velocity $\velscalb$, ranging in $\{ 0, 0.25, 1/\sqrt{2}, 1, 2\}$, with the associated colours ranging from red to yellow. Each panel corresponds to a different value of $\ac$, as indicated at the top, showing a large range of behaviours for the translational dynamics depending on $\ac$ and $\velscalb$. In particular, the top row features negative values of $\ac$, for which the effective velocity $\veleff$ vanishes when $\velscalb = -\velscala/\ac$; this makes the associated trajectories near invisible on the figure, because they essentially remain at the origin. In each panel, we use $\Breth = 0.5$, $\omb = 10$, $\velscala = 1$, $\velscalc = 0$,  initial position $\Xvecpos = \vec{0}$, and initial orientation $(\theta,\psi,\phi) = (2\pi/5,\pi/2,-7\pi/15)$, simulating over the same time interval.}
\label{fig:translational_GD_2}
\end{figure}

The conditional significance of $\velscalb$ is illustrated in Figures \ref{fig:translational_GD},  \ref{fig:translational_GD_2}, and Supplementary Movie 2, which explore the emergent translational dynamics more generally. Figures \ref{fig:translational_GD}a and \ref{fig:translational_GD}b highlight how $\velscalc$ has no effect on the leading-order translation, with the average trajectories in each panel being qualitatively indistinct, while we see an overall trend that increasing $\ac$ decreases the effective swimming speed when $\velscalb=0$, as predicted by our analytic result \eqref{fin0}. In Figures \ref{fig:translational_GD}c and \ref{fig:translational_GD}d, however, we see that increasing $\velscalb$ from zero does modify the swimming trajectory, with these modifications increasing in significance as $\abs{\ac}$ increases, in line with the predictions of \eqref{fin0}. We emphasise that $\veleff \to \velscalb$ as $\abs{\ac} \to \infty$, in line with physical intuition, despite the apparent importance of the sign of $\ac$. This is because $\etilde{1}(\alp,\phic)$ depends on $\ac$ and, in fact, tends to exact opposite directions in the limits $\ac \to \pm \infty$ (when the initial conditions derived in Appendix \ref{sec: Slow time functions IC} are taken into account). Figure \ref{fig:translational_GD_2} further emphasises the significant differences that can be observed on the translational dynamics upon switching the sign of $\ac$.

\section{Results and conclusions}\label{sec: results}
In this section, we summarise the main results and conclusions that can be drawn from our detailed asymptotic analysis.

\subsection{Dynamics in the bacterial limit}
\label{sec: summary bacteiral limit}
In the `bacterial' limit, fast rotation occurs only about the spheroidal axis of symmetry. In \S \ref{sec: bacterial sublimit: angular dynamics}, which corresponds to the limit $\ac \to 0$ in the general case considered in \S \ref{sec: General emergent behaviour}, we showed that rapid rotation in the bacterial limit does not materially impact on the emergent angular dynamics. In particular, the leading-order dynamics of all but the spin angle $\psi$ are exactly as in the rotation-free case, so that rapid rotation leaves the direction of the swimmer symmetry axis unchanged. Further, if we consider the leading-order \emph{average} dynamics, then the spin angle evolves precisely as if there were no rapid rotation. Hence, the average orientation of the swimmer is not altered by rapid spinning about the symmetry axis, and hence one can safely neglect rapid rotation without consequence when considering leading-order angular dynamics in the bacterial limit.

However, one must be careful with the effect of rapid spinning in this limit when concerned with the \emph{translational} dynamics of the swimmer. In \S \ref{sec: bacterial sublimit: translational dynamics}, we showed that the leading-order dynamics of self propulsion are dominated by the component of velocity along the axis of symmetry. Physically, this is because the other components of linear velocity cancel out over a period of the fast axial rotation, so one can safely neglect off-axis components of translation at leading order. Moreover, the effective direction of self propulsion follows the average orientation of the symmetry axis, which evolves as if there were no rapid spinning.

Thus, overall, our analysis demonstrates that a spheroid that spins rapidly about its symmetry axis can be reliably modelled as a non-spinning version of the same spheroid in a shear flow. Moreover, we have shown that any relatively slow, off-axis rotation can be neglected, as can any components of translational velocity that are not aligned with the axis of symmetry. In the context of the idealised bacterial swimmers that motivated this limit, our analysis supports the use of simple, effective models of bacterial swimmers that do not include any details of off-axis rotation or off-axis translation, which might plausibly arise due to the finite length of a propulsive flagellum or fascicle, for instance.

\subsection{Dynamics for general rapid spinning}
For general rapid spinning, the axis of rotation does not coincide with the symmetry axis, and the straightforward simplifications of the bacterial limit no longer hold. Nevertheless, from our analysis in \S \ref{sec: General emergent behaviour} for $\ac = \order{1}$, we are still able to characterise and quantify the emergent behaviours that arise from the multiscale dynamics, generalising those seen in the bacterial case. For a more intuitive interpretation of our results it is helpful to work with $\angl$, the angle between the rotational and symmetry axis, instead of $\ac$. These quantities are related through (the principal branch of) $\tan \angl = \ac$.

\begin{figure}
    \centering
    \medskip
    \begin{overpic}[width=0.51\textwidth]{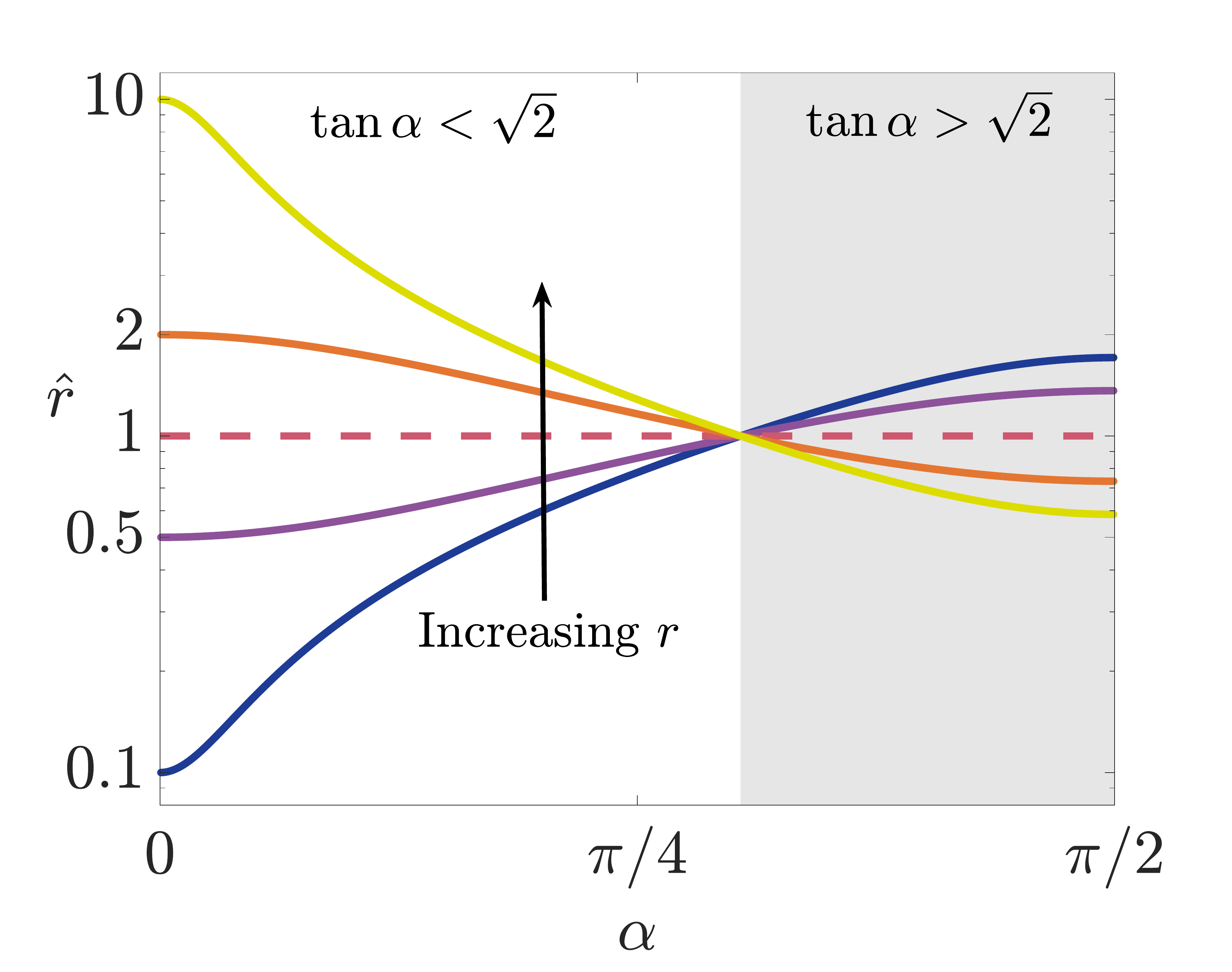}
    \put(0,70){(a)}
    \end{overpic}
    \hspace{-4ex}
    \begin{overpic}[width=0.51\textwidth]{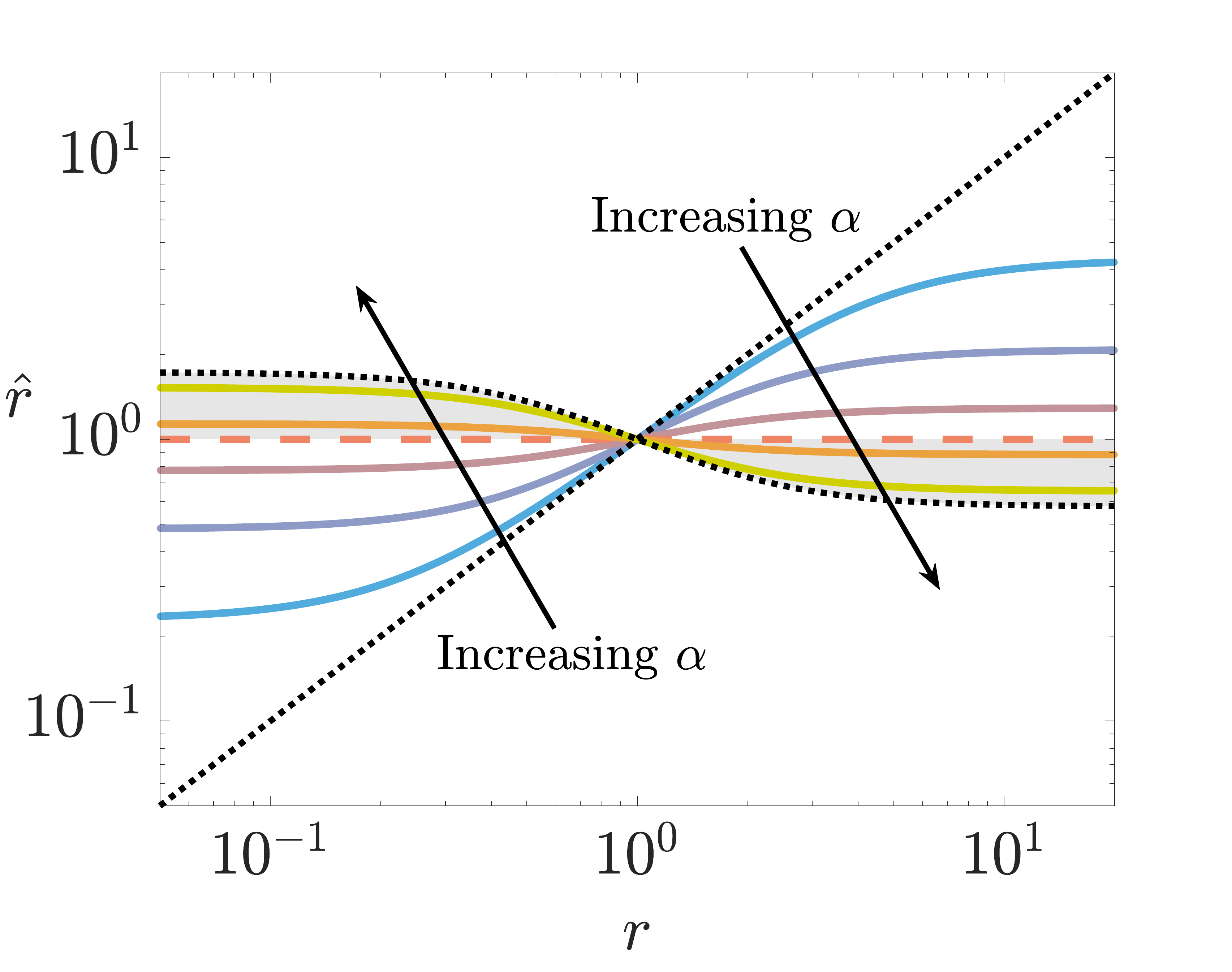}
    \put(2,70){(b)}
    \end{overpic}
    \caption{The effective aspect ratio $\ratioeff$ from \eqref{eq: effective ratio} as a function of (a) $\angl$, the angle between the rotational and symmetry axes, and (b) $\ratio$, the spheroid aspect ratio. The shaded area in both subfigures denotes where spinning prolate spheroids appear as effective (passive) oblate spheroids and vice-versa. In (a), the different curves correspond to $\ratio \in \left\{0.1, 0.5, 1, 2, 10 \right\}$; hence, $\ratioeff = \ratio$ when $\angl = 0$, in accordance with the results of \S \ref{sec: Bacterial sublimit}. Additionally, the dashed line denotes the special case $\ratio = 1$, from which we see that the effective shape of a rotating sphere is always a sphere i.e. $\ratioeff \equiv 1$. In (b), the different lines correspond to $\angl \in \left\{0, \pi/12,\pi/6, \pi/4, \arctan \sqrt{2}, \pi/3, 5\pi/12, \pi/2  \right\}$. The black dotted lines denote the special cases $\angl = 0, \pi/2$ and the dashed line denotes the special angle $\angl = \arctan \sqrt{2}$, for which all rotating spheroids behave as effective spheres.}
\label{fig:Effective_aspect_ratio}
\end{figure}

A key result of our general analysis is that rapid spinning of spheroidal objects always leads to effective Jeffery's orbits, though now in transformed variables that can be associated with the original angular variables. In addition, we have analytically calculated the parameter that characterises these orbits: the effective Bretherton parameter $\Beff$, defined in \eqref{eq: effective coefficients}, which depends nonlinearly on $\angl$. This effective parameter $\Beff$ can be sufficiently different from the original Bretherton parameter $\Breth = (\ratio^2 - 1)/(\ratio^2 + 1)$ (where $\ratio$ is the length-to-diameter ratio of the spheroid) so as to completely alter the character of the orbit. Associating the effective Bretherton parameter $\Beff$ with an effective aspect ratio $\ratioeff$ via $\Beff = (\ratioeff^2 - 1)/(\ratioeff^2 + 1)$, and using our key result for $\Beff$ in \eqref{eq: effective coefficients}, we may deduce an explicit expression for the effective aspect ratio:
\begin{align}
\label{eq: effective ratio}
\ratioeff = \sqrt{\dfrac{4 \ratio^2 \cos^2 \angl + \left(3 + \ratio^2\right) \sin^2 \angl}{4 \cos^2 \angl + \left(1 + 3 \ratio^2 \right) \sin^2 \angl}}.
\end{align}
We illustrate the effective aspect ratio $\ratioeff$ in Figure \ref{fig:Effective_aspect_ratio} as a function of $\angl$ and $\ratio$, noting that $\ratioeff$ is even in $\angl$ (as one might expect intuitively). Our results show that rapid rotation modifies the effective hydrodynamic shape of the spheroid through its effective aspect ratio and the natural parameterisation of its orientation. In Supplementary Movie 3, we show how the average rotational dynamics of a rapidly spinning spheroid are well represented by the rotational dynamics of a passive spheroid with effective Bretherton parameter \eqref{eq: effective coefficients} (which corresponds to an aspect ratio \eqref{eq: effective ratio}).

We see that $\ratioeff = \ratio$ for $\angl = 0$, in agreement with the results of \S\ref{sec: Bacterial sublimit} summarised in \S\ref{sec: summary bacteiral limit}. As $\angl$ increases, the effective aspect ratio $\ratioeff$ changes monotonically, with the effective shape of prolate spheroids ($\ratio > 1$) becoming less prolate and eventually oblate, and vice versa for oblate spheroids ($\ratio < 1$). Thus, the distortion of the effective shape is most pronounced when $\angl = \pi/2$ i.e. when the rotation and symmetry axes are perpendicular. In this scenario, we see from \eqref{eq: effective ratio} that $\ratioeff^2 = (3 + \ratio^2)/(1 + 3 \ratio^2)$. In particular, this tells us that an extremely prolate spheroid rapidly rotating along an axis perpendicular to its symmetry axis will behave as an oblate spheroid with effective aspect ratio $1/\sqrt{3}$, no matter the original aspect ratio of the spheroid. Similarly, an extremely oblate spheroid will behave as a prolate spheroid with effective ratio $\sqrt{3}$. Hence, the effective hydrodynamic shape is not simply the envelope of the rotating shape.

A particularly striking result of our analysis is the scenario in which $\tan \angl = \sqrt{2}$, which leads to an effective aspect ratio of $\ratioeff = 1$ for all values of the original aspect ratio $\ratio$. That is, if the rotational axis is at an angle $\angl = \arctan \sqrt{2} \approx \SI{54.7}{\degree}$ from the symmetry axis of a spheroid, the spinning spheroid will behave hydrodynamically as an effective sphere in the shear flow. This universal reduction can be seen in Figure \ref{fig:Effective_aspect_ratio}. It can also be seen in \Cref{fig:rotational_GD}d, with the orbits of differently shaped spheroids collapsing onto the same dynamics for $\ac = \sqrt{2}$ (i.e. $\angl = \arctan \sqrt{2}$). This identification of an active spheroidal particle behaving like a sphere in three-dimensional shear flow complements recent work in which a passive non-spheroidal particle (a symmetric boomerang with angle $\arccos (2 \sqrt{3} - 3) \approx \SI{62.3}{\degree}$) was shown to behave like a sphere in two-dimensional shear flow \citep{roggeveen2022motion}.

Now considering the translational dynamics, we again see that effective propulsion occurs along the average symmetry axis. However, in this general case, from \eqref{fin0} the effective speed of propulsion is $\veleff = \velscala \cos \angl + \velscalb \sin \angl$. This gives an intuitive explanation for the effective speed of propulsion; it is exactly the magnitude of the projection of $\vel$ along the rotational axis $\angvel$. That is, $\veleff = (\vel \bcdot \angvel)/|\angvel|$. This means that the effective speed is now influenced by both the on-axis speed $\velscala$ and the component of propulsion in the direction of off-axis spinning $\velscalb$, with their relative contributions modulated by their relative contributions along the rotational axis. In addition, as $\angl \rightarrow \pi/2$ the off-axis component of propulsion $\velscalb$ dominates the on-axis propulsion $\velscala$ (though the maximal speed of self-propulsion is always bounded above by $\sqrt{\velscala^2 + \velscalb^2}$). Hence, in contrast to the bacterial limit, off-axis propulsion cannot simply be neglected for spheroidal objects undergoing off-axis spinning; here, the effects of propulsion parallel to the direction of off-axis spinning can combine with the off-axis spinning itself to contribute to the effective propulsion.

Indeed, the relative signs of the translational and angular components of propulsion can significantly alter the emergent behaviour. For example, if $\velscala\approx-\velscalb$, then spinning with $\angl = \pi/4$ can lead to propulsion having essentially no leading-order contribution. Increasing $\angl$ beyond this value highlights that the direction of effective propulsion can strongly depend on $\angl$ and, therefore, on the relative direction of the rotational axis in comparison to the symmetry axis. In particular, it is possible to be in a scenario where the reversal of precisely one of the components of rotation gives rise to both a reversal and a change of magnitude of the effective propulsive velocity.

\section{Discussion}
\label{sec: Discussion}

In this first of two studies, we have explored the behaviours of rapidly spinning spheroidal objects in three-dimensional Stokes shear flow. Using the method of multiple scales for systems, we have derived effective governing equations that, when written in terms of appropriately transformed variables, are simply the classic Jeffery's equations \citep{Jeffery1922,Bretherton1962}, with appropriately modified Bretherton parameter $\Beff$. This result is in a similar spirit to \citet{Bretherton1962} and \citet{Brenner1964a}, who showed that Jeffery's equations hold for the motion of general axisymmetric objects, with the Bretherton parameter being interpreted as an effective aspect ratio. Indeed, our work shows that Jeffery's equations also hold for active axisymmetric (not just spheroidal) objects. That is, the rotational dynamics of rapidly spinning axisymmetric objects also behave as effective passive spheroids in shear flow, with a
corresponding effective aspect ratio that depends on the particular object, and is \emph{not} just the envelope of the spinning object. Through our analysis, we analytically calculate how the rapid spinning modifies this effective parameter in the effective equations, and hence the effective aspect ratio (and therefore shape) of the spheroid. This identification of modified, effective parameterisations of the original 3D governing equations is similar to those found in other recent applications of multi-timescale analysis to planar swimming problems \citep{Walker2022,Walker2022a}, suggesting that such reductions may be commonplace in similarly posed problems of Stokes flow.

In two opposite limits of the rapid spinning problem, where spinning is dominated by either the on- or off-axis component, we obtain substantial simplifications to our general results. The case where spinning about the symmetry axis is dominant can be considered as a simplified model of a rotating bacterial swimmer.\footnote{Acknowledging that much of the detail of the complex swimming problem has been neglected in this idealised study of rigid bodies.} In this bacterial limit, our results show that one can effectively ignore any off-axis propulsion and all components of rapid spinning, so that the swimmer can be effectively modelled as a non-spinning object with axial propulsion. In the other limit, where the dominant spinning is perpendicular to the symmetry axis, while the angular dynamics now exhibit strong (leading-order) oscillations, the long-term angular dynamics are still governed by effective Jeffery’s equations in terms of appropriate transformed variables. Moreover, the effective propulsion is still along the `average' symmetry axis, as in the bacterial case, though the effective propulsive velocity is now dominated by the off-axis component of instantaneous propulsion, perhaps counterintuitively.

In Part 2 of this two-part study, we will consider swimmers of a more general shape. Specifically, we will study helicoidal swimmers with chirality, generalising our results beyond the spheroidal swimmers considered in Part 1. As one might expect, this additional generality complicates the analysis we have presented here. Nevertheless, we will see that it remains analytically tractable via the techniques employed in this work. 

\textbf{Acknowledgements.} M.P.D. is supported by the UK Engineering and Physical Sciences Research Council [Grant No. EP/W032317/1]. C.M. is a JSPS Postdoctoral Fellow (P22023) and acknowledges support by the JSPS-KAKENHI Grant-in Aid for JSPS Fellows (Grant No. 22F22023). K.I. acknowledges JSPS-KAKENHI for Young Researchers (Grant No. 18K13456), JSPS-KAKENHI for Transformative Research Areas (Grant No. 21H05309), JST, PRESTO (Grant No. JPMJPR1921) and JST, FOREST (Grant No. JPMJFR212N). B.J.W. is supported by the Royal Commission for the Exhibition of 1851. 

\textbf{Declaration of interests.} The authors report no conflict of interest.

\textbf{Data accessibility.} Minimal computer code for exploring the dynamics, as well as the scripts used to generate the figures in this study are available at \url{https://github.com/Clementmoreau/spinningswimmers}.

\appendix

\setcounter{equation}{0}
\renewcommand{\theequation}{\thesection\arabic{equation}}

\section{Defining the Euler angles}
\label{sec: deriving eqs of motion}

\begin{figure}
\vspace*{0.5cm} 
    \centering
    \includegraphics[width=0.9\textwidth]{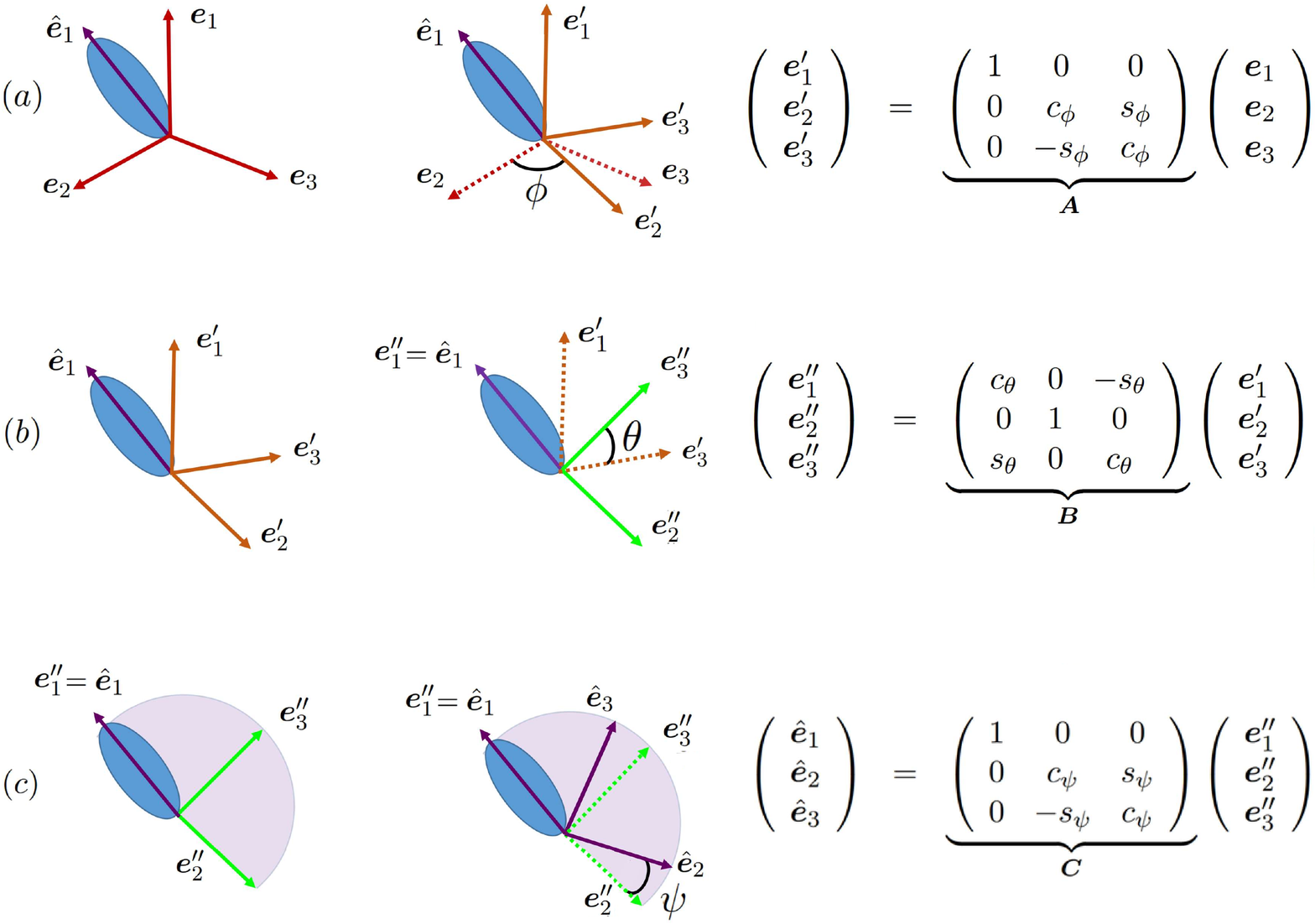}
\vspace*{0.5cm} 
 \caption{The three rotations generating the $xyx$-Euler transformation from the laboratory fixed frame basis, $\{\e{1}, \e{2}, \e{3}\}$ to the swimmer fixed frame basis, $\{\ehat{1}, \ehat{2}, \ehat{3}\}$ for a spheroidal swimmer, as schematically represented by the ellipse in the figure.  ($a$) The first rotation is of angle $\phi$ about the laboratory frame basis vector $\e{1}$, with the basis vectors before and after the rotation related by the rotation matrix $\bm A$, as given in the figure. ($b$)  The second rotation is of angle $\theta$ about the intermediate basis vector $\eprime{2}$, which is the same as the intermediate basis vector $\eprimeprime{2}$,  with the intermediate basis vectors $\{\eprime{1}, \eprime{2}, \eprime{3}\}$  related to the basis $\{\eprimeprime{1}, \eprimeprime{2}, \eprimeprime{3}\}$ by the rotation matrix $\bm B$. ($c$)  The third and final  rotation is of angle $\psi$ about the intermediate basis vector $\ehat{1} =\eprimeprime{1}$, with the intermediate basis  $\{\eprimeprime{1}, \eprimeprime{2}, \eprimeprime{3}\}$ related to the swimmer fixed frame basis $\{\ehat{1}, \ehat{2}, \ehat{3}\}$ by the rotation matrix $\bm C$.}
    \label{fig: euler angles}
\end{figure}

Here, we derive the Euler angles used in the main text. We relate the bases of the laboratory frame and the swimmer frame by an $xyx$-Euler angle transformation. In particular, with $c_\theta$, $s_\theta$ denoting $\cos\theta$, $\sin\theta$, and similarly for other angles, we have
\begin{align}
    \eihatmat = \mat{C}\mat{B}\mat{A} \eimat 
    = \left(\begin{array}{c|c|c}
    c_\theta & s_\phi s_\theta & - c_\phi s_\theta \\ \label{elm2} 
    s_\psi s_\theta &  \hphantom{+}c_\phi c_\psi  - s_\phi c_\theta s_\psi  &  \hphantom{+}s_\phi c_\psi  + c_\phi c_\theta s_\psi \\
   c_\psi s_\theta &  - c_\phi s_\psi  -s_\phi c_\theta c_\psi  & 
   -s_\phi s_\psi  + c_\phi c_\theta c_\psi
   \end{array}\right) \eimat , 
\end{align}
as illustrated in \Cref{fig: euler angles}. Here, $\mat{A}$ is the matrix for a rotation of angle $\phi$ about $\e{1}$, $\mat{B}$ is the matrix for a rotation of angle $\theta$ about the $y$-axis of the resulting reference frame, with associated basis vector 
\begin{equation}
    \eprime{2}=\eprimeprime{2}=\cos\psi \ehat{2}-\sin\psi\ehat{3},    
\end{equation}
and $\mat{C}$ is the matrix for a rotation of angle $\psi$ about the body fixed axis, $\ehat{1}$. Intermediate basis vectors are defined in \Cref{fig: euler angles}.

The inverse of this mapping is generated by applying the rotations in reverse order with negated angles. Thus, by applying the transformation $(\phi,\theta,\psi)\mapsto(-\psi,-\theta,-\phi)$ to the matrix in \eqref{elm2}, we have
\begin{equation}   
   \eimat = \left(
 \begin{array}{c|c|c} 
   c_\theta & s_\psi s_\theta & c_\psi s_\theta \\
   \hphantom{+}s_\phi s_\theta & c_\phi c_\psi - s_\phi c_\theta s_\psi  & -c_\phi s_\psi  - s_\phi c_\theta c_\psi \\
   -c_\phi s_\theta & s_\phi c_\psi  + c_\phi c_\theta s_\psi  & -s_\phi s_\psi  + c_\phi c_\theta c_\psi 
   \end{array}\right) \eihatmat. \label{elm1}
\end{equation} 
Further, the Euler angle transformation also gives the relation between the angular velocity of the swimmer frame in the presence of flow, denoted $\angvel^f$, and the time derivatives of the Euler angles via
\begin{equation}
    \angvel^{f} = \dot{\phi}\e{1} + \dot{\theta}\eprime{2} + \dot{\psi}\ehat{1} = \sum \hat{\Omega}_p^f \ehat{p},
\end{equation}
which simplifies to 
\begin{equation}\label{ang1} 
    \begin{pmatrix}
        \dot{\theta}\\
        \dot{\psi}\\
        \dot{\phi}\\
    \end{pmatrix} 
    = 
    \begin{pmatrix}
        0 &  c_\psi  & -s_\psi \\
        1 & -s_\psi c_\theta/s_\theta & -c_\psi c_\theta/s_\theta    \\
        0 &  \hphantom{+}s_\psi/ s_\theta & c_\psi/s_\theta          
    \end{pmatrix}
    \begin{pmatrix}
        \hat{\Omega}^{f}_1 \\
        \hat{\Omega}^{f}_2\\
        \hat{\Omega}^{f}_3
    \end{pmatrix}.
\end{equation}

\section{Helical trajectories in the absence of shear flow}
\label{sec: helix}

In this appendix, we discuss the helical trajectory of a spinning microswimmer in the absence of any background flow. We rederive the well-known helix theorem of \citet{shapere1989geometry}, demonstrating that a swimmer with constant swimming velocity $\bm{V}$ and angular velocity $\bm{\Omega}$ follows a helical trajectory. In particular, we provide a relation between the swimming velocity and the properties of the helix, such as its radius and pitch.

In the absence of any background flow, there is a single timescale present in the angular dynamics, so that they are in fact captured exactly by \eqref{eq: full gov eq trans LO General}, interpreted as ordinary differential equations in time $T$ and with $\alp$, $\muc$, and $\phic$ being constants. For clarity, we denote the solutions to these helical angular dynamics by $\hel{\theta}$, $\hel{\psi}$, and $\hel{\phi}$. Since there is no background flow, without loss of generality we may set the initial orientation of the swimmer to be $\hel{\theta}(0)=\pi/2$, $\hel{\phi}(0)=\arctan(-1/\ac)$, and  $\hel{\psi}(0)=0$ for later convenience, and we may take $\ehat{1}$ to be parallel to the swimming velocity, so that $\vel = V\ehat{1}$, where $V$ is the swimming speed.

From the analysis of \S \ref{sec: leading-order}, the now-constant $\alp$ obeys \eqref{eq: alpha def}. Given our initial conditions, we therefore have $\cos\alp=0$, so that $\alp=\pm\pi/2$. Substituting this into \eqref{eq: om sin thet sq def} gives the time evolution of $\hel{\theta}$ as
\begin{align}
    \om \cos\hel{\theta}=\mp\ac \cos(T+\muc),
\end{align}
and we obtain the second constant of integration as $\muc=\pm\pi/2$ from the initial conditions. The choice of sign in $\alp$ and $\muc$ is determined by the initial values of the derivative in \eqref{eq: full gov eq trans LO General}, giving $\alp = -\pi/2$, yielding
\begin{align}
\label{eq: helix theta}
    \cos\hel{\theta}=-\frac{\ac}{\om}\sin T.
\end{align}
The dynamics of $\hel{\psi}$ are then obtained from \eqref{eq: sin theta sin psi} and \eqref{eq: sin theta cos psi} as 
\begin{align}
\label{eq: helix psi}
    \tan\hel{\psi}=\frac{1}{\om}\tan T.
\end{align}
The third and final constant of integration $\phic$ is obtained using \eqref{eq: phi0 sol}, which here reads
\begin{align}
    \tan(\hel{\phi}-\phic)=\ac \cos T.
\end{align}
The initial condition for $\hel{\phi}$ then gives $\phic=- \pi/2$, yielding the dynamics of $\hel{\phi}$ as
\begin{align}
\label{eq: helix phi}
    \tan\hel{\phi}= -\frac{1}{\ac \cos T}.
\end{align}

With reference to the Euler angle parameterisation given explicitly in \eqref{elm1}, the translational dynamics $\mathrm{d}\Xvecpos/\mathrm{d}t=V\ehat{1}$ can be written in the lab frame as
\begin{align}
 \om\frac{\mathrm{d}X}{\mathrm{d}T}=V\cos\hel{\theta}\,,\quad
 \om\frac{\mathrm{d}Y}{\mathrm{d}T}=V\sin\hel{\phi}\sin\hel{\theta}\,,\quad
 \om\frac{\mathrm{d}Z}{\mathrm{d}T}=-V\cos\hel{\phi}\sin\hel{\theta}\,.
\end{align}
By substituting equations \eqref{eq: helix theta}, \eqref{eq: helix psi}, and \eqref{eq: helix phi} into these translational evolution equations gives, after integration in time,
\begin{align}
    X= \frac{\ac}{\om^2}V(\cos T-1)\,,\quad
    Y=-\frac{1}{\om^2}VT\,,\quad
    Z=-\frac{\ac}{\om^2}V\sin T\,,
\end{align}
which describe a helical trajectory. The radius $\rho$ and pitch $b$ of the helix are then given by
\begin{align}
    \rho=\frac{\ac}{\om^2}V, \qquad 
    b=\frac{2\pi}{\om^2}V.
\end{align}

\section{Initial conditions for the slow-time functions}
\label{sec: Slow time functions IC}

In this Appendix, we provide the appropriate initial conditions for the slow-time functions i.e. $\alp(0)$, $\muc(0)$, and $\phic(0)$, given generic initial conditions for the original variables i.e. $\theta(0)$, $\psi(0)$, and $\phi(0)$.

We first use \eqref{eq: alpha def} to determine that
\begin{align}
\label{eq: alpha def IC}
\om \cos \alp(0) = \ac \sin \theta_0(0) \sin \psi_0(0) +  \cos \theta_0(0).
\end{align}
We then combine (\ref{eq: mu def}a,c,d) to deduce that
\begin{align}
\label{eq: muc def IC}
\tan \muc(0) = \dfrac{\om \sin \theta(0) \cos \psi(0)}{\ac \cos \theta(0) - \sin \theta(0) \sin \psi(0)}.
\end{align}
Since $\cos$ and $\tan$ are not monotonic, \eqref{eq: alpha def IC} and \eqref{eq: muc def IC} do not uniquely prescribe $\alp(0)$ and $\muc(0)$. Therefore, we must prescribe additional consistency conditions using results from \S \ref{sec: General emergent behaviour}. To this end, we generate the following two additional consistency conditions from (\ref{eq: mu def}a,c,d):
\begin{subequations}
\label{eq: consistency check first}
\begin{align}
\sin \alp(0) \sin \muc(0) &= -\sin \theta(0) \cos \psi(0), \\
\om \sin \alp(0) \cos \muc(0) &= \sin \theta(0) \sin \psi(0) - \ac \cos \theta(0).
\end{align}
\end{subequations}
Then, we perform the following procedure to determine $\alp(0)$ and $\muc(0)$. We take the principle values of $\arccos$ and $\arctan$ to calculate $\alp(0) \in [0, \pi]$ from \eqref{eq: alpha def IC}, and to generate a potential value of $\muc(0) \in [-\pi/2, \pi/2]$ from \eqref{eq: muc def IC}. We then check whether this potential value of $\muc(0)$ is consistent with \eqref{eq: consistency check first}. If it is not, we are on the wrong branch of $\muc$, and so we instead use $\muc(0) \mapsto \muc(0) + \pi$, which means $\muc(0) \in [\pi/2, 3\pi/2]$.

Finally, we must determine $\phic(0)$. To do this, we use \eqref{eq: phi0 sol} to deduce that
\begin{align}
\label{eq: phi0 sol IC}
\tan(\phi(0) - \phic(0)) = \dfrac{\ac \sin \muc(0)}{\ac \cos \alp(0) \cos \muc(0) + \sin \alp(0)}.
\end{align}
Taking the principle value of $\arctan$ gives $\phic(0) \in [\phi(0) -\pi/2, \phi(0) + \pi/2]$. To ensure we are on the correct branch, we then impose a consistency condition using \eqref{eq: a 0 LO identities General a sin thet cos phi}, namely:
\begin{align}
\label{eq: a 0 LO identities General a sin thet cos phi IC}
\om \sin \theta(0) \cos \phi(0) = \cos \phic(0) \left(\ac \cos \alp(0) \cos \muc(0) + \sin \alp(0) \right) - \ac \sin \phic(0) \sin \muc(0).
\end{align}
If $\phic(0)$ is not consistent with \eqref{eq: a 0 LO identities General a sin thet cos phi IC}, we are on the wrong branch of $\phic$, and so we instead use $\phic(0) \mapsto \phic(0) + \pi$, which means $\phic(0) \in [\phi(0) + \pi/2, \phi(0) + 3\pi/2]$.

\section{Solving the next-order adjoint problem}
\label{sec: next order adjoint}
In this Appendix, we solve the 3D homogeneous adjoint problem $L^* \bsX = 0$ from \eqref{eq: adjoint system} for $\bsX = (\Xb, \Xc, \Xa)^{T}$, where the adjoint operator $L^*$ is defined in \eqref{eq: adjoint matrix}.

We first note that the last row in \eqref{eq: adjoint matrix} generates the straightforward solution
\begin{align}
\label{eq: Z1 and A1}
\Xa(\ts) = \Ca,
\end{align}
where $\Ca$ is an arbitrary constant. This simplifies the system \eqref{eq: adjoint system}, \eqref{eq: adjoint matrix} into the following 2D system
\begin{subequations}
\label{eq: Z2 Z3 orig}
\begin{align}
\om \dbyd{\Xb}{\ts} + \ac \Xc \dfrac{\sin \psi_0}{\sin^2 \theta_0} &= \ac \Ca \dfrac{\cos \theta_0 \sin \psi_0}{\sin^2 \theta_0} , \\
\om \dbyd{\Xc}{\ts} - \ac \Xb \sin \psi_0  - \ac \Xc \dfrac{\cos \theta_0 \cos \psi_0}{\sin \theta_0} &= - \ac \Ca \dfrac{\cos \psi_0}{\sin \theta_0}.
\end{align}
\end{subequations}
Then, it is straightforward to note that the particular integral
\begin{align}
\label{eq: Z2/Z3 and A1}
\Xb(\ts) = 0, \quad \Xc(\ts) = \Ca \cos \theta_0,
\end{align}
addresses the right-hand side of \eqref{eq: Z2 Z3 orig}, using the leading-order \cref{eq: theta eq trans LO General} to evaluate the derivative of $\cos \theta_0$. Thus, \eqref{eq: Z1 and A1}, \eqref{eq: Z2/Z3 and A1} give one linearly independent solution to the adjoint problem \eqref{eq: adjoint system}, in terms of the arbitrary constant $\Ca$.

To obtain the remaining two linearly independent solutions, it is helpful to subtract off the particular integral solution we have already obtained, using the mapping $\Xc(\ts) \mapsto \Ca \cos \theta_0 + \Xc(\ts)$, transforming \eqref{eq: Z2 Z3 orig} into the system
\begin{subequations}
\label{eq: Z2 Z3 homog}
\begin{align}
\om \dbyd{\Xb}{\ts} + \ac \Xc \dfrac{\sin \psi_0}{\sin^2 \theta_0} &= 0, \\
\om \dbyd{\Xc}{\ts} - \ac \Xb \sin \psi_0  - \ac  \Xc \dfrac{\cos \theta_0 \cos \psi_0}{\sin \theta_0} &= 0.
\end{align}
\end{subequations}
To obtain solutions to \eqref{eq: Z2 Z3 homog}, we are motivated by the knowledge that
\begin{align}
\label{eq: gen homog sol to thet_1 and ps_1}
\theta_1  = \om \pbyp{\theta_0}{\ts} = \ac \cos \psi_0, \qquad \psi_1  = \om \pbyp{\psi_0}{\ts} = 1 - \ac \dfrac{\sin \psi_0 \cos \theta_0}{\sin \theta_0},
\end{align}
are solutions to the homogeneous versions of the original linear problem \eqref{eq: O eps system}, \eqref{eq: L and F def} (the factors of $\om$ in \eqref{eq: gen homog sol to thet_1 and ps_1} are for algebraic convenience). This follows from noting that the homogeneous version of \eqref{eq: O eps system}, \eqref{eq: L and F def} is equivalent to differentiating \eqref{eq: theta eq trans LO General}--\eqref{eq: psi eq trans LO General} with respect to $\ts$. Therefore, if we can find substitutions that transform \eqref{eq: Z2 Z3 homog} into the homogeneous versions of the original linear problem \eqref{eq: O eps system}, \eqref{eq: L and F def}, then we obtain solutions using \eqref{eq: gen homog sol to thet_1 and ps_1}.

The first such transformation is $\Xb(\ts) = -\xbs (\ts) \sin \theta_0$, $\Xc(\ts) = \xcs (\ts) \sin \theta_0$, which maps \eqref{eq: Z2 Z3 homog} into
\begin{subequations}
\label{eq: Z2 Z3 homog subst}
\begin{align}
\om \dbyd{\xbs}{\ts} - \ac \xcs \dfrac{\sin \psi_0}{\sin^2 \theta_0} + \ac \xbs \dfrac{\cos \theta_0 \cos \psi_0}{\sin \theta_0} &= 0, \\
\om \dbyd{\xcs}{\ts} + \ac \xbs \sin \psi_0 &= 0.
\end{align}
\end{subequations}
Setting $\xbs = \psi_1$ and $\xcs = \theta_1$, \eqref{eq: Z2 Z3 homog subst} is equivalent to the homogeneous version of \eqref{eq: O eps system}, \eqref{eq: L and F def}. Noting the result \eqref{eq: gen homog sol to thet_1 and ps_1}, this means that the second linearly independent solution of the adjoint problem \eqref{eq: adjoint system} is
\begin{align}
\label{eq: Z2/Z3 and A3}
\Xa(\ts) = 0, \quad \Xb(\ts) = \Cb \left(\ac \sin \psi_0 \cos \theta_0 - \sin \theta_0 \right), \quad \Xc(\ts) = \ac \Cb \sin \theta_0 \cos \psi_0,
\end{align}
where $\Cb$ is an arbitrary constant.

The second such transformation is $\Xc(\ts) \mapsto \Xc (\ts) \sin^2 \theta_0$, which maps \eqref{eq: Z2 Z3 homog} directly into the homogeneous version of \eqref{eq: O eps system}, \eqref{eq: L and F def} with $\Xb = \theta_1$ and $\Xc = \psi_1$. Therefore, using the result \eqref{eq: gen homog sol to thet_1 and ps_1}, the third and final linearly independent solution to the adjoint problem \eqref{eq: adjoint system} is
\begin{align}
\label{eq: Z2/Z3 and A2}
\Xa(\ts) = 0, \quad \Xb(\ts) = \ac \Cc \cos \psi_0, \quad \Xc(\ts) = \Cc \sin \theta_0 \left( \sin \theta_0 - \ac \sin \psi_0 \cos \theta_0 \right),
\end{align}
where $\Cc$ is an arbitrary constant.

Hence, the three linearly independent solutions to the 3D system \eqref{eq: adjoint system}, \eqref{eq: adjoint matrix} are described by \eqref{eq: Z1 and A1}, \eqref{eq: Z2/Z3 and A1}, \eqref{eq: Z2/Z3 and A3}, and \eqref{eq: Z2/Z3 and A2}. Moreover, we note that these solutions can also be verified by direct substitution into the system \eqref{eq: adjoint system}, \eqref{eq: adjoint matrix}.

\section{Evaluating the left-hand sides of the solvability conditions \eqref{eq: solv conditions orig}}
\label{sec: LHS of solv cond}

In this Appendix, we evaluate the left-hand sides of \eqref{eq: solv conditions orig}, currently written in terms of the original variables $\theta_0(\ts,\tl)$, $\psi_0(\ts,\tl)$, and $\phi_0(\ts,\tl)$. To proceed, we need to write them in terms of the slow-time functions $\alp(\tl)$, $\muc(\tl)$, and $\phic(\tl)$.

We start with the most straightforward simplification, the left-hand side of \eqref{eq: solv conditions orig fb fc 1}. By taking the derivative of \eqref{eq: alpha def} with respect to $\tl$, we obtain the relationship
\begin{align}
\label{eq: om sin alpha da/dt}
\theta_{0 \tl} \left(\ac \cos \theta_0 \sin \psi_0 - \sin \theta_0 \right) + \psi_{0 \tl} \ac \sin \theta_0 \cos \psi_0 = - \om \sin \alp \dbyd{\alp}{\tl},
\end{align}
which coincides with the integrand in the left-hand side of \eqref{eq: solv conditions orig fb fc 1}. This agreement is not unexpected, since we expect one of our solvability conditions to involve the slow-time variation of the fast-time conserved quantity \eqref{eq: alpha def}. Since the right-hand side of \eqref{eq: om sin alpha da/dt} is independent of $\ts$, it is straightforward to take its fast-time average, yielding the following simplification of the left-hand side of \eqref{eq: solv conditions orig fb fc 1}:
\begin{align}
\label{eq: solv cond alp}
\av{\theta_{0 \tl} \left(\ac \cos \theta_0 \sin \psi_0 - \sin \theta_0 \right) + \psi_{0 \tl} \ac \sin \theta_0 \cos \psi_0} = - \om \sin \alp \dbyd{\alp}{\tl}.
\end{align}

The next simplest case is the left-hand side of \eqref{eq: solv conditions orig fb fc 2}. To simplify this, we note that the integrand of the left-hand side of \eqref{eq: solv conditions orig fb fc 2} can be evaluated through the sums of the following identities:
\begin{subequations}
\label{eq: general fb/fc identities}
\begin{align}
\label{eq: theta_0t}
\theta_{0 \tl} \ac \cos \psi_0 - \psi_{0 \tl} \ac \cos \theta_0 \sin \theta_0 \sin \psi_0 &= \ac \cos^2 \theta_0 \pbyp{}{\tl} \left(\dfrac{\sin \theta_0 \cos \psi_0}{\cos \theta_0} \right), \\
\label{eq: ps_0t}
\psi_{0 \tl} \sin^2 \theta_0 &=  -\sin^2 \psi_0 \sin^2 \theta_0 \pbyp{}{\tl} \left( \cot \psi_0 \right) ,
\end{align}
\end{subequations}
which follow from directly evaluating the derivatives on the right-hand sides. To effectively exploit the right-hand sides of \eqref{eq: ps_0t}, we require the relationship
\begin{align}
\label{eq: cot psi}
\cot \psi_0 &= -\dfrac{\om \sin \alp \sin (\ts + \muc)}{\sin \alp \cos (\ts + \muc) + \ac \cos \alp},
\end{align}
which follows from taking the ratio of \eqref{eq: sin theta cos psi} and \eqref{eq: sin theta sin psi}. Then, substituting the relationships \eqref{eq: mu def}, \eqref{eq: cot psi} into the right-hand sides of \eqref{eq: general fb/fc identities}, we can sum the two subequations and rearrange to deduce that
\begin{align}
\label{eq: first fb/fc identity}
\theta_{0 \tl} \ac \cos \psi_0 + \psi_{0 \tl} \sin \theta_0 \left(\sin \theta_0 - \ac  \cos \theta_0 \sin \psi_0 \right) = \om \sin^2 \alp \dbyd{\muc}{\tl}.
\end{align}
Since the right-hand side of \eqref{eq: first fb/fc identity} is independent of $\ts$, it is straightforward to take its fast-time average. This yields the following simplification of the left-hand side of \eqref{eq: solv conditions orig fb fc 2}:
\begin{align}
\label{eq: solv cond mu}
\av{\theta_{0 \tl} \ac \cos \psi_0 + \psi_{0 \tl} \sin \theta_0 \left(\sin \theta_0 - \ac \cos \theta_0 \sin \psi_0 \right)} = \om \sin^2 \alp \dbyd{\muc}{\tl}.
\end{align}

Our final task for this subsection is to deal with the left-hand side of \eqref{eq: solv conditions orig fa fc}. The first step is to differentiate the arctangent of \eqref{eq: phi0 sol} with respect to $\tl$ to deduce
\begin{align}
\pbyp{\phi_0}{\tl} &= \dbyd{\phic}{\tl} +\ac \dbyd{\alp}{\tl}\dfrac{ \sin (\ts + \muc) \left[\ac \sin \alp \cos (\ts + \muc)  -  \cos \alp \right]}{(\ac \cos \alp \cos (\ts + \muc)  + \sin \alp)^2 + \ac^2 \sin^2 (\ts + \muc) } \notag \\
\label{eq: phi 0t}
&\quad + \ac \dbyd{\muc}{\tl} \dfrac{\ac \cos \alp + \sin \alp \cos (\ts + \muc) }{(\ac \cos \alp \cos (\ts + \muc)  + \sin \alp)^2 + \ac^2 \sin^2 (\ts + \muc)}.
\end{align}
Then, we combine \eqref{eq: ps_0t} and \eqref{eq: phi 0t}, substitute in the relationships \eqref{eq: mu def} and rearrange to deduce that
\begin{align}
\label{eq: fa/fc identity}
\psi_{0\tl} \cos \theta_0 + \phi_{0\tl} = \cos \alp \dbyd{\muc}{\tl} + \dbyd{\phic}{\tl}.
\end{align}
Similarly to before, the right-hand side of \eqref{eq: fa/fc identity} is independent of the fast-time $\ts$, so it is straightforward to take its fast-time average. Hence, we find that
\begin{align}
\label{eq: solv cond phic}
\av{\psi_{0\tl} \cos \theta_0 + \phi_{0\tl}} = \cos \alp \dbyd{\muc}{\tl} + \dbyd{\phic}{\tl}.
\end{align}

This concludes our evaluation of the left-hand sides of \eqref{eq: solv conditions orig} in terms of derivatives of the slow-time functions $\alp$, $\muc$, and $\phic$. These derived relationships are given in \eqref{eq: solv cond alp}, \eqref{eq: solv cond mu}, and \eqref{eq: solv cond phic}, and allow us to re-write \eqref{eq: solv conditions orig} as \eqref{eq: solv conditions transformed}.

\section{Fast-time average of the swimmer basis}
\label{sec: fast time average of swimmer basis}
In this Appendix, we evaluate the fast-time average of the swimmer-fixed basis functions, defined in terms of the angular variables and laboratory-fixed basis in \eqref{elm2}. This allows us to calculate the fast-time average of the right-hand side of \eqref{eq: transform gov eq translational FC}.

To determine the fast-time average of $\ehat{1}(\theta_0,\phi_0)$, defined in \eqref{elm2}, we simply calculate the fast-time averages of the trigonometric expressions \eqref{eq: om cos thet def}, \eqref{eq: a 0 LO identities General a} as follows:
\begin{subequations}
\label{eq: averages for translation 1}
\begin{align}
    \om \e{1} \bcdot \ehat{1} (\theta_0,\phi_0) &= \cos \alp - \ac \sin \alp \cos \sig, \\
    \om \e{2} \bcdot \ehat{1} (\theta_0,\phi_0) &= \ac \cos \phic \sin \sig + \sin \phic \left(\ac \cos \alp \cos \sig + \sin \alp \right), \\
    \om \e{3} \bcdot \ehat{1} (\theta_0,\phi_0) &=  \ac \sin \phic \sin \sig -\cos \phic \left(\ac \cos \alp \cos \sig + \sin \alp \right). 
\end{align}
\end{subequations}
Taking the fast-time average of the relationships \eqref{eq: averages for translation 1}, the fast-time average of $\ehat{1}(\theta_0,\phi_0)$ is
\begin{align}
    \av{\om \ehat{1}(\theta_0,\phi_0)} = \cos \alp  \e{1} +\sin \alp \sin \phic \e{2} - \sin \alp \cos\phic \e{3} = \etilde{1}(\alp,\phic),
\end{align} 
where $\etilde{1}(\alp,\phic)$ can be considered equivalent to the (hatted) basis vector $\ehat{1}$ in \eqref{elm2}, but with argument $(\theta,\phi)$ replaced by $(\alp,\phic)$.

To determine the fast-time average of $\ehat{2}(\theta_0,\psi_0,\phi_0)$, defined in \eqref{elm2}, we note that 
\begin{subequations}
\label{eq: averages for translation 2}
\begin{align} 
    \om \e{1} \bcdot \ehat{2} (\theta_0,\psi_0,\phi_0) &= \ac \cos \alp + \sin \alp \cos \sig, \\
    \om \e{2} \bcdot \ehat{2} (\theta_0,\psi_0,\phi_0) &= \ac \sin \alp \sin \phic - \cos \phic \sin \sig - \cos \alp \sin \phic \cos \sig, \\
    \om \e{3} \bcdot \ehat{2} (\theta_0,\psi_0,\phi_0) &= \cos \alp \cos \phic \cos \sig - \ac \sin \alp \cos \phic - \sin \phic \sin \sig,
\end{align}
\end{subequations}
using the expressions \eqref{eq: mu def}, \eqref{eq: a 0 LO identities General a}. Taking the fast-time average of the relationships \eqref{eq: averages for translation 2}, the fast-time average of $\ehat{2}(\theta_0,\psi_0,\phi_0)$ is
\begin{equation}
    \av{\om \ehat{2}(\theta_0,\psi_0,\phi_0)} = \ac \left( \cos \alp  \e{1} + \sin \alp \sin \phic \e{2} -\sin \alp \cos\phic \e{3} \right) = \ac \etilde{1}(\alp,\phic).
\end{equation}

Finally, to determine the fast-time average of $\ehat{3}(\theta_0,\psi_0,\phi_0)$, defined in \eqref{elm1}, we note that  
\begin{subequations}
\label{eq: averages for translation 3}
\begin{align}
\e{1} \bcdot \ehat{3} (\theta_0,\psi_0,\phi_0) &= - \sin \alp \sin \sig, \\
\e{2} \bcdot \ehat{3} (\theta_0,\psi_0,\phi_0) &= \cos \alp \sin \phic \sin \sig - \cos \phic \cos \sig, \\
\e{3} \bcdot \ehat{3} (\theta_0,\psi_0,\phi_0) &= - \cos \alp \cos \phic \sin \sig - \sin \phic \cos \sig, 
\end{align}
\end{subequations}
using \eqref{eq: sin theta sin psi}. Noting that the fast-time average of each relationship in \eqref{eq: averages for translation 3} vanishes, the fast-time average of $\ehat{3}(\theta_0,\psi_0,\phi_0)$ is
\begin{align}
\av{\om \ehat{3}(\theta_0,\phi_0)} = \vec{0}.
\end{align}

\bibliographystyle{jfm_draft}
\bibliography{reference.bib,BJWMendeley.bib}

\end{document}